\documentclass[useAMS,usenatbib]{mn2e}
\usepackage{graphicx}
\usepackage{amssymb}

\title[The prompt afterglow connection in Gamma-Ray Burst]{The prompt-afterglow connection in Gamma-Ray Bursts: a comprehensive
statistical analysis of Swift X-ray light-curves}
\author[R. Margutti et al.]{R. Margutti$^{1,2}$ \thanks{E-mail:
rmargutti@cfa.harvard.edu (RM)}, E. Zaninoni$^{2,3}$, M.~G. {Bernardini}$^{2}$, G. {Chincarini}$^{4,2}$, F. Pasotti$^{2}$, \and
C. Guidorzi$^{5}$, L. Angelini$^{6}$, D.~N. Burrows$^{7}$, M. Capalbi$^{8}$, P.~A. Evans$^{9}$, N. Gehrels$^{6}$, \and
 J. Kennea$^{7}$, V. Mangano$^{10}$, A. Moretti$^{2}$, J. Nousek$^{7}$, J.~P. Osborne$^{9}$, K.~L. Page$^{9}$,  \and
 M. Perri$^{8}$,
 J. Racusin$^{6}$, P. Romano$^{10}$, B. Sbarufatti$^{2,7}$, S. Stafford$^{11}$, M. Stamatikos$^{11,6}$ \\
$^{1}$ Harvard-Smithsonian Center for Astrophysics, 60 Garden Street, Cambridge, MA02138\\
$^{2}$ INAF Osservatorio Astronomico di Brera, via Bianchi 46, Merate 23807, Italy \\
$^{3}$ University of Padova, Physics \& Astronomy Dept.  Galileo Galilei via Marzolo, 8 - I-35131 Padova\\
$^{4}$ Univerisit\'a Milano Bicocca, Dip. Fisica G. Occhialini, P.zza della Scienza 3, Milano 20126, Italy\\ 
$^{5}$ Dipartimento di Fisica, Univerisit\'a di Ferrara, via Saragat 1, I-44122 Ferrara, Italy\\
$^{6}$ NASA-Goddard Space Flight Center, Greenbelt, MD, USA \\
$^{7}$ Department of Astronomy \&  Astrophysics, 525 Davey Lab, USA\\
$^{8}$ ASI Science Data Center, via G. Galilei, 00044 Frascati, Italy\\
$^{9}$ X-ray and Observational Astronomy Group, Department of Physics \& Astronomy, University of Leicester, LE1 7RH, UK \\
$^{10}$ INAF-IASF Palermo, Via Ugo La Malfa 153, Palermo, Italy \\
$^{11}$ Dept. of Physics and Center for Cosmology and Astro-Particle Physics, Ohio State University, Columbus, OH 43210, USA}
\begin{document}

\date{Accepted Year Month Day. Received Year Month Day; in original form Year Month Day}
\pagerange{\pageref{firstpage}--\pageref{lastpage}} \pubyear{2011}
\maketitle
\label{firstpage}

\begin{abstract}
We present a comprehensive statistical analysis of Swift X-ray light-curves of Gamma-Ray Bursts (GRBs) 
collecting data from more than 650 GRBs discovered by Swift and other facilities.
The unprecedented sample size allows us to constrain the \emph{rest-frame}
X-ray properties of GRBs from a statistical perspective, with particular reference to intrinsic time scales and the 
energetics of the different light-curve phases in a common rest-frame 0.3-30 keV energy band.
Temporal variability episodes 
are also studied and their properties constrained. 
Two fundamental questions drive this effort: i) Does the X-ray emission retain any kind of ``memory''
of the prompt $\gamma$-ray phase? ii) Where is the dividing line between long and short GRB X-ray properties?
We show that short GRBs decay faster, are less luminous and less energetic than long GRBs in the X-rays,
but  are interestingly characterized by similar intrinsic absorption. We furthermore reveal the 
existence of a number of statistically significant relations that link the X-ray to prompt $\gamma$-ray
parameters in long GRBs;  short GRBs are outliers of the majority of these 2-parameter relations. However and more
importantly, we report on the existence of a universal 3-parameter scaling that links the 
X-ray and the $\gamma$-ray energy to the prompt spectral peak energy of \emph{both}
long and short GRBs: $E_{\rm{X,iso}}\propto E_{\rm{\gamma,iso}}^{1.00\pm 0.06}/E_{\rm{pk}}^{0.60\pm 0.10}$.
\end{abstract}

\begin{keywords}
gamma-ray: bursts -- radiation mechanism: non-thermal --X-rays
\end{keywords}
\section{Introduction}

In 7 years of operation, \emph{Swift} \citep{Gehrels04} has revolutionized our understanding of the 
X-ray emission that follows the prompt $\gamma$-ray phase of  Gamma-Ray Bursts (GRBs): the X-ray 
afterglow.
The standard afterglow theory (\citealt{Meszaros97}; \citealt{Sari98}) predicts that X-ray emission arises from the interaction of 
a relativistic outflow with the ambient medium, leading to the formation of a blast wave. 
In this context, short and long GRBs would naturally show similar afterglows (since 
the emission would be sensitive to the energy budget of the relativistic outflow but otherwise keep no memory 
of the origin of the outflow) \emph{if} the properties of the environment of the two classes are also similar.
This is difficult to reconcile with the massive star (long GRBs, see e.g. \citealt{Woosley93}) vs. compact binary (short GRBs
see e.g. \citealt{Paczynski86}) 
progenitor systems, which would instead suggest a wind density profile (i.e. $\propto r^{-2}$)
around long GRBs and an ISM (i.e. $\propto r^{0}$) ambient density for short bursts.
Contrary to expectations, observations are often consistent with a constant density environment (ISM)
also in the case of long GRBs (e.g. \citealt{Racusin09}). 
Moreover, the standard afterglow theory fails to explain the presence of long ($\sim10^4$ s)
phases characterized by very mild decays (the so-called plateaus or shallow decay phases)
in the X-ray light-curve of many GRBs (e.g. GRB\,060729, \citealt{Grupe07}); it cannot account for abrupt drops of emission
observed in some GRBs (e.g. GRB\,070110, \citealt{Troja07}, \citealt{Lyons10}) and has serious difficulties explaining the X-ray flares
(\citealt{Chincarini07}, \citealt{Falcone07}; \citealt{Chincarini10} and references therein). 

As a result, a number of alternative models have been proposed. They basically divide into
two classes: accretion onto a newly born black hole which directly powers the observed X-ray 
light-curve \citep{Kumar08} or power from a rapidly rotating magnetar (see e.g. \citealt{Metzger11}
and references therein).
In sharp contrast to the standard afterglow theory, those models directly relate the properties
of the observed X-ray light-curves to the GRB central engine. 

With this work we improve our understanding of the X-ray emission of long and short 
GRBs through a homogeneous analysis of a sample of more than 650 GRBs observed by
 \emph{Swift}-XRT \citep{Burrows05}.
We ask: what is the typical amount of energy released during the different X-ray light-curve phases?
Does the X-ray emission retain any kind of ``memory" of the prompt $\gamma$-ray phase?
GRBs are traditionally classified into long and short according to their prompt $\gamma$-ray
properties: do short GRBs show a distinct behaviour in the X-rays as well?
Is it possible to find a \emph{universal} (i.e. common to long and short GRBs) scaling that involves
prompt and X-ray properties? 
This set of still open questions constitutes the major reason to undertake the present investigation.

Previous attempts mainly concentrated on \emph{observer} frame properties and tried
to understand  to what extent the observations could be reconciled with the standard forward shock model
(e.g. \citealt{Obrien06,Butler07b,Butler07c,Willingale07,Liang07,Liang08,Evans09,Racusin09,Racusin11}).
This effort led to the identification of serious difficulties of the standard picture.
We build on previous results and adopt here a different approach: instead of 
comparing observations with a particular physical model,
we take advantage of the large sample size and look for
\emph{correlations} between the X-ray and $\gamma$-ray properties of GRBs \emph{any}
physical model will have to explain. We complement previous studies with: 
\begin{enumerate}
\item[i.] Homogeneous analysis of GRBs in a \emph{common} \emph{rest frame} energy band (0.3-30 keV).
\item[ii.] Statistics and properties of the temporal variability superimposed on the smooth X-ray decay.
\item[iii.] Comparative study of  long vs. short GRB X-ray afterglows.
\item[iv.] Study of the prompt $\gamma$-ray vs.  X-ray connection: notably, we report on the existence of 
a \emph{universal} scaling involving prompt and X-ray parameters.
\end{enumerate} 

Hereafter,  we will refer to the X-ray signal recorded by \emph{Swift}-XRT
after a GRB trigger as ``X-ray light-curve" (LC) and explicitly do not use the word ``afterglow" 
to avoid confusion (``afterglow" refers to the standard interpretation). 
Uncertainties are given at 68\% confidence
level (c.l.) unless explicitly mentioned. Standard cosmological quantities have been adopted: 
$H_{0}=70\,\rm{km\,s^{-1}\,Mpc^{-1}}$, $\Omega_{\Lambda }=0.7$, $\Omega_{\rm{M}}=0.3$.
The results from our analysis are publicly available.\footnote{A demo
version of the website is currently available at 
http://www.grbtac.org/xrt\_demo/GRB060312Afterglow.html }
\section{Data analysis}
\label{Sec:dataanalysis}

We select GRBs observed by \emph{Swift}-XRT from the beginning of science operations in
December 2004 through the end of 2010. The starting sample includes 658 GRBs; 36 belong to 
the class of  short GRBs. Following \cite{Margutti11b}, the short or long nature of each event 
is established using the combined information 
from the duration, hardness and spectral lag of its prompt	 $\gamma$-ray emission:	a prompt $\gamma$-ray 
duration	$T_{90}\lesssim2$ s coupled to a hard $\gamma$-ray emission with photon index $\Gamma \lesssim 1.5$
and a negligible  spectral lag are considered indicative of a short GRB nature. 

For each GRB, the data reduction comprises four steps:
\begin{itemize}
\item Extraction of count-rate light-curves (LC hereafter) in the 0.3-10 keV energy band.
\item Time-resolved spectral analysis.
\item Flux calibration in the observer frame 0.3-10 keV energy band and luminosity calibration in the
rest-frame 0.3-30 keV energy range for the sub-sample of GRBs with known redshift.
\item LC fitting.
\end{itemize}

The \emph{Swift}-XRT data have been analyzed using the latest version of the HEASOFT package available at the time of 
the analysis (v. 6.10). For each GRB we started from calibrated event lists and sky images as distributed by the
HEASARC archive\footnote{http://heasarc.gsfc.nasa.gov/cgi-bin/W3Browse/swift.pl}. 
The following analysis made extensive use of the XRTDAS software package. The additional automated 
processing was performed via custom IDL scripts. Details on the procedure followed
can be found in \cite{MarguttiPhD}.\footnote{Retrievable from http://hdl.handle.net/10281/7465} 
Here we note that: 
\begin{itemize}
\item[(i)] While agreeing on the major steps, the data extraction adopted 
here is slightly different from the methods presented in \cite{Evans07} and \cite{Evans09}: however,
after comparing every single LC obtained with the two techniques, we find that the methods lead to
consistent results. 
\item[(ii)] Our flux and luminosity LC calibration is based on a time-resolved spectral analysis which is
able to capture the spectral evolution of the source with time. Uncertainties arising from the
spectral analysis have also been propagated into the final flux and luminosity LCs (this is essential
to compute  the significance of positive temporal fluctuations superimposed on the smoothly decaying LC,
see Sec. \ref{SubSec:fitting}).
\item[(iii)] The intrinsic neutral hydrogen absorbing column $\rm{NH_{HG}}$ of each GRB was computed 
extracting a spectrum in the widest interval of time with no apparent spectral evolution. 
The best-fitting $\rm{NH_{HG}}$ was used as frozen parameter in the time-resolved spectral
analysis above. The Galactic contribution was frozen to the value in the direction of the burst 
as computed by \cite{Kalberla05}.
\end{itemize}

Since we corrected for the Galactic and intrinsic absorption,
the final results are \emph{unabsorbed} 0.3-10 keV (observer frame) flux LCs. For the sub-sample of
GRBs with known redshift we furthermore extracted \emph{unabsorbed} luminosity LCs in the
0.3-30 keV (rest-frame) energy band (extrapolating the best-fitting power-law spectrum). 
We conservatively use only $z$ derived from optical spectroscopy and photometric redshifts for
which potential sources of degeneracy (e.g. dust extinction) can be ruled out with high confidence.
The complete list of redshifts used is reported in Table \ref{Table:bestfitpar} (175 GRBs in our sample
have redshift).
\subsection{Light-curve fitting}
\label{SubSec:fitting}

\begin{figure}
    \includegraphics[scale=0.5]{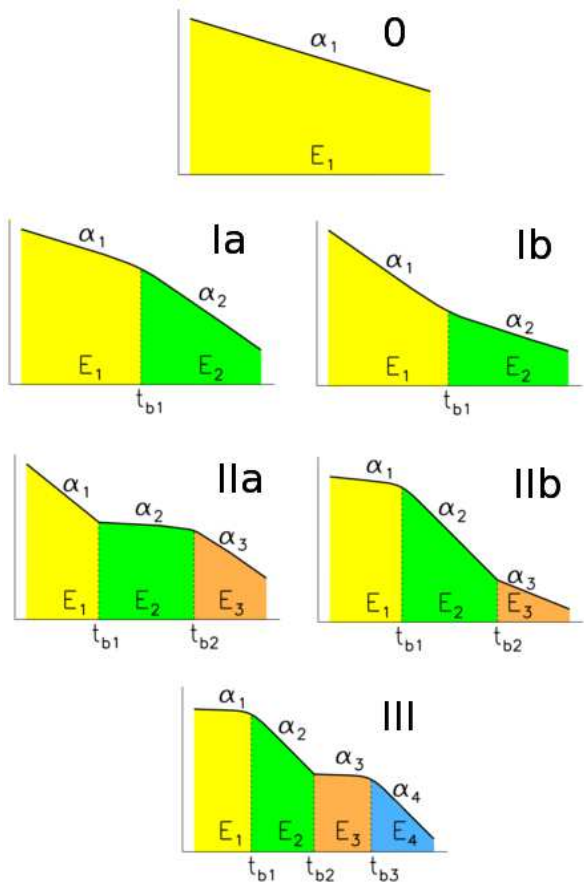}
\caption{Cartoon showing the different LC models used in this work. Both axes use logarithmic units.}
\label{Fig:lctype}
\end{figure}

\begin{figure}
\includegraphics[scale=0.4]{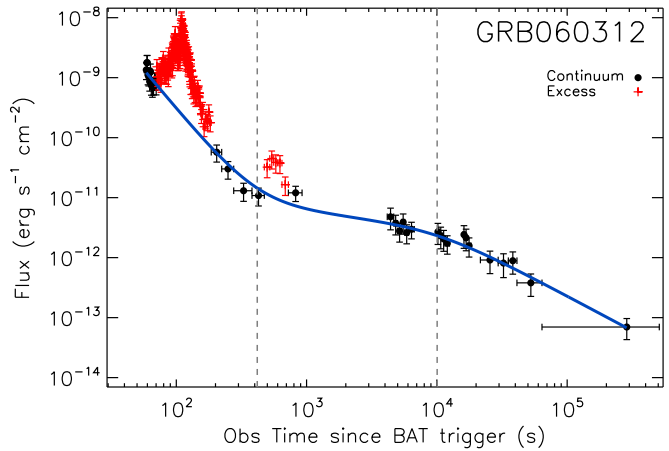}
\caption{0.3-10 keV unabsorbed flux LC of GRB\,060312 with best-fitting model superimposed (blue solid line).
The red crosses mark the data points identified as ``excesses" during the semi-automatic fitting 
routine and subsequently removed from the fit to obtain the best-fitting power-law plus broken power-law model.
Vertical dashed lines: best estimates of the break times obtained as described in the main text.}
\label{Fig:lcfit}
\end{figure}

\begin{table}
\centering
\caption{Number of GRBs per LC type. For each type, the number of GRBs with complete LCs
(C-GRBs, defined as promptly repointed GRBs $t_{\rm{rep}}<300$ s whose fading was followed up down to a factor
$\sim 5-10$ from the background limit) and with detected positive fluctuations (F-GRBs) with respect to 
the best-fitting smooth decay is also reported. GRBs are classified as either C or U-like, and either F or N-like. U-like bursts have truncated LCs, while
N-like GRBs show no evidence for flares. This classification refers to the 0.3-10 keV (observer frame)
LCs.}
\label{Tab:lctype}
\begin{tabular}{|c|cccccc|}

\hline
 &  \multicolumn{6}{|c|}{Light-curve types}  \\
\hline
 &0 & Ia  & Ib & IIa & IIb & III\\
 \hline
 Total Number of GRBs  & 114 & 89 & 61 & 133 & 18 & 22\\
\hline
C-GRBs                         & 42  & 61 & 53 & 121 & 17 & 22\\
\hline
F-GRBs                          & 23 & 16 & 24 & 48    & 8  & 10 \\
\hline
\end{tabular}
\end{table}

The X-ray LCs of GRBs for $t\gtrsim60\,\rm{s}$ consist of smoothly decaying power-laws or broken power-laws
with X-ray flares superimposed.
Here we concentrate on the underlying smooth component.

We considered only GRBs whose statistics were good enough to allow us to extract a spectrum to 
convert their count-rate LCs into flux LCs (total of 437 GRBs out of 658). We first fitted the
entire sample of flux LCs in the 0.3-10 keV (observer frame) energy band.
We then focussed on the sub-sample of GRBs with redshift and performed a second fit using the 
LCs in the common 0.3-30 keV (rest-frame).

Our semi-automatic fitting routine is based on the $\chi^2$ statistic and closely 
follows the procedure outlined in \cite{Margutti11a}. We fit the following models.
Defining: 
\begin{equation}
	f(N_1,\alpha_1,t)\equiv N_1\,t^{-\alpha_1}
\end{equation}
(where $N_1$ and $\alpha_1$ are the normalization and the slope of the power-law, respectively)  and
\begin{equation}
	g(N_2,\alpha_2,\alpha_3,t_{b},s,t)\equiv N_2\left( \left( \frac{t}{t_{b}}\right)^{-\frac{\alpha_2}{s}}+ \left(\frac{t}{t_{b}}\right)^{-\frac{\alpha_3}{s}} \right)^{s}
\end{equation}
(where $N_2$ is the normalization, $\alpha_2$ and $\alpha_3$ are the slopes of the broken power-law; $t_{b}$ is the break time while  $s$ is the smoothing
parameter), the fitting models can be written as: 
\begin{itemize}
	\item Simple power-law (model \textbf{0}):
		\begin{eqnarray}
		F=f(N_1,\alpha_1,t) \label{Eq:first}
		\end{eqnarray}
	\item Smoothed broken power-law (model \textbf{Ia} and \textbf{Ib}  for $s<0$ and $s>0$, respectively):
		\begin{eqnarray}
		F=g(N_1,\alpha_1,\alpha_2,t_{b1},s_1,t)
		\end{eqnarray}
	\item Smoothed broken power-law plus initial (model \textbf{IIa}) power-law decay:
		\begin{eqnarray}
		F=f(N_1,\alpha_1,t)+g(N_2,\alpha_2,\alpha_3,t_{b2},s_1,t)
		\end{eqnarray}
		or final (model \textbf{IIb}) power-law decay:
		\begin{eqnarray}
		F=g(N_1,\alpha_1,\alpha_2,t_{b1},s_1,t)+f(N_2,\alpha_3,t)
		\end{eqnarray}
	\item	Double  smoothly joined broken power-laws (model \textbf{III}):
		\begin{eqnarray}
		F=g(N_1,\alpha_1,\alpha_2,t_{b1},s_1,t)+g(N_2,\alpha_3,\alpha_4,t_{b3},s_2,t) \label{Eq:last}
		\end{eqnarray}
\end{itemize}
The model number follows from the number of break times.  For model \textbf{IIa} (\textbf{IIb}, \textbf{III})
the first (second) break time is defined as the time when the second (first) component outshines the first (second)
component. Figure \ref{Fig:lctype} illustrates the different models, while Fig. 
 \ref{Fig:lcfit} shows the result for GRB\,060312 taken as an example: 
in this case the semi-automatic procedure identified two episodes of emission in excess of the smooth decay
(model \textbf{IIa}).
The number of GRBs for each LC type are listed in Table \ref{Tab:lctype}. We refer to the LCs as
``type" 0, Ia, Ib, IIa, IIb and type III  GRBs in the following.

The best-fitting parameters together with their uncertainties and associated covariance matrix were then used
to derive the 0.3-10 keV (observer-frame) fluence 
of the entire LC, from the \emph{Swift}-XRT re-pointing time to the end of the 
observation. No temporal extrapolation was applied at this stage. Note that the contribution from
significant positive fluctuations has \emph{not} been included. The fluence of the different LC phases as defined by
the temporal breaks was also calculated (Fig. \ref{Fig:lctype}). Results are listed in Table \ref{Table:bestfitpar} and
Table \ref{Table:bestfitene}.
We then followed the very same procedure to fit the 0.3-30 keV (rest-frame) LCs:
Table \ref{Table:bestfitene0330} reports the energetics in this energy range.

The list of LC points flagged as ``excesses"  during the fitting procedure (e.g. red crosses of Fig. \ref{Fig:lcfit}) 
constituted for each GRB the starting sample to look for significant positive fluctuations with respect to the best fit. 
The information contained in the covariance matrix was used to derive the uncertainties associated with the 
residuals with respect to the best-fit (residuals were at this stage calculated on the \emph{entire} LC). 
We first selected positive fluctuations with a minimum $1\sigma$ significance. We furthermore require the
positive fluctuations to show a rise plus decay \emph{structure}: this procedure automatically excluded single 
data-points scattering from the best fit. GRBs showing (not showing) such structures were flagged as ``F"  (``N")
in Table \ref{Table:bestfitpar}-\ref{Table:bestfitene0330}.  
GRB\,060312 in Fig. \ref{Fig:lcfit}, with two rising and decaying structures 
superimposed on the smooth decay qualifies as ``F"-event. The fluence (energy for known $z$) of those excesses
was calculated by simply integrating the flux of each LC bin over the bin duration (after subtracting the
contribution from the underlying smoothly decaying emission). Errors were propagated
accordingly and can be used to quantify how significant is the presence of emission in addition to the
smooth power-law decay in each GRB. The fluence (energy) of positive fluctuations detected during the different
LC phases (e.g. steep-decay, plateau, normal-decay, etc.) was also derived and listed in Table 
\ref{Table:bestfitene}  (0.3-10 keV, observer frame) and Table \ref{Table:bestfitene0330} (0.3-30 keV, rest frame). 
For simplicity, in the following we will use the word ``flare" to refer to statistically significant
positive fluctuations detected on top of the smoothly decaying component, being however aware that
different kinds of variability possibly contribute to the detected ``flaring activity".
\section{Long vs. Short GRBs properties}
\label{Sec:results}

\begin{table}
\label{Tab:SGRBs}
\centering
\begin{tabular}{l}
\hline
Short GRBs\\
\hline
\textbf{050724} 051221A \textbf{051227} 060313 \textbf{061006}\\ 
061201 \textbf{070714B} 070724A 070809 \textbf{071227}\\ 
080123 \textbf{080503} 080919 090510 090515\\ 
090607 100117A 100816A 101219A\\
\hline
\end{tabular}
\caption{List of 19 short GRBs with complete LCs. GRBs with detected 
temporally extended emission
are in boldface \citep{Norris11}.}
\end{table}

The analysis above reduces the X-ray LC of GRBs to a set of measured 
parameters:  temporal slopes; break times; total isotropic energy
(fluence) and energy (fluence) associated  with the different LC phases;  flare energy (fluence); 
spectral photon index temporal evolution; intrinsic neutral hydrogen absorption. 
This constitutes an unprecedented set of information
homogeneously obtained on the largest sample of GRBs to date and represents the natural sample to 
look for correlations among the parameters.

While we report the
best-fitting parameters of the entire sample (Tables \ref{Table:bestfitpar}-\ref{Table:bestfitene0330}), 
in the following we restrict our analysis
to GRBs with ``complete" LCs, defined as those GRBs re-pointed by XRT at  $t_{\rm{rep}}< 300$ s and
for which we were able to follow the fading of the XRT flux down to a factor $\sim 5-10$ from the background
limit (or, equivalently, $t_{\rm{end}}\gtrsim 4\times 10^5$ s). These GRBs are flagged as ``C" in Tables 
\ref{Table:bestfitpar}-\ref{Table:bestfitene0330}.
The number of ``C"-like GRBs per LC morphological type is reported in Table \ref{Tab:lctype}.
``U"-like GRBs have instead truncated LCs. Short GRBs with complete LCs are listed in 
Table \ref{Tab:SGRBs}. Short GRBs with extended emission are in boldface
(see \citealt{Norris11}).
\subsection{Median X-ray light-curve of long and short GRBs}
\label{SubSec:aftdis}

\begin{figure}
\includegraphics[width=1.05\hsize,clip]{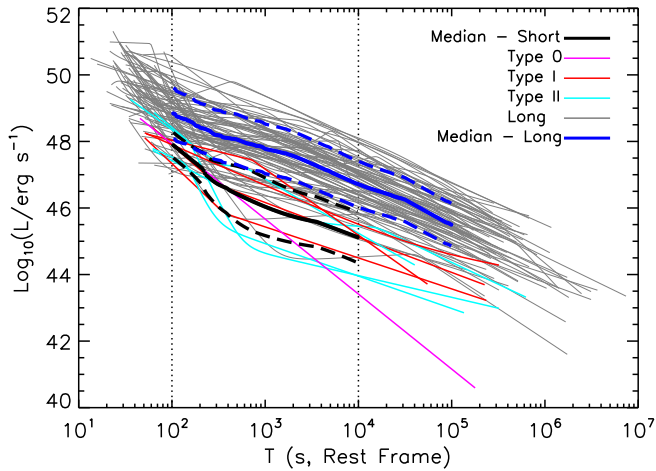}
\caption{Best fitting profiles of
9 C-like short GRBs with rest-frame time coverage $10^2-10^4$s (colour lines) superimposed on the sample of long GRBs (grey lines).
Black (blue) solid line:  median $Log(L)$ light-curve for short (long) GRBs; dashed lines mark the $1\sigma$ dispersion.
The LCs have been calibrated in the rest-frame 0.3-30 keV energy band.}
 \label{Fig:dispersion}
\end{figure}

We select the sub-sample of  C-like LCs of GRBs with redshift
observed in the common rest-frame time interval $10^2-10^5$ s and $10^2-10^4$ s for long and short GRBs, 
respectively. These criteria resulted in a sample of 79 long GRBs and 9 short GRBs  (Fig. \ref{Fig:dispersion}). 
We consider here 0.3-30 keV (rest-frame) luminosity LCs.

We combined the best-fitting profiles of Sec. \ref{SubSec:fitting} to produce a median luminosity LC
of a long GRB. The result is shown in Fig. \ref{Fig:dispersion}:
the median luminosity LC roughly decays as $\sim t^{-1}$ (with a milder decay $\sim t^{-0.9}$
for $(0.5<t<4)\,\rm{ks}$ and a steeper $\sim t^{-1.2}$ decay after $\sim 4\,\rm{ks}$).
With the possible exception of the shallower section, this is in rough agreement with the
prediction of the standard afterglow theory\footnote{A steepening
to $\sim1.5-2$ is predicted after the jet-break time if the outflow is collimated into a 
jet (\citealt{Rhoads99}; \citealt{Sari99}).} (\citealt{Meszaros97}; \citealt{Sari98}).

The decay of the median LC is steeper for short ($\propto t^{-1.3}$ ) than for long GRBs 
($\propto t^{-1}$) in the rest-frame time interval $10^2-10^4$ s;  short GRB X-ray LCs
are on average less luminous by a factor $\sim 10-30$ than long GRBs X-ray LCs.
This conclusion holds also considering long GRBs in the \emph{same} redshift bin.
However, Fig. \ref{Fig:dispersion} clearly shows that the two samples slightly overlap
(see also \citealt{Gehrels08}). The steeper 
decay that characterizes short GRBs causes a progressive shift of their luminosity distribution
towards the low end of the long GRB distribution.

\begin{table*}
\centering
\caption{Characteristic quantities describing the parameter distributions (number of elements ($\#$), mean ($m$), median ($M$), standard deviation ($SD$), skewness ($SK$)) and best-fitting values from a Gaussian fit  (mean ($\mu$), standard deviation ($\sigma$), normalization ($N$)). Fluences (S) are given in $10^{-6}$ erg cm$^{-2}$, energies (E) in $10^{50}$ erg, fluxes (F) in $10^{-6}$ erg s$^{-1}$ cm$^{-2}$, luminosities (L) in $10^{48}\,\rm{erg\,s^{-1}}$, times (t) in s, hydrogen column densities (NH) in $10^{22}$ cm$^{-2}$. Note that logarithmic (linear) units have been used in the upper (lower) half of the Table. 
We refer the reader to Appendix \ref{Appendix:par} for the exact definition of the parameters listed below. X-ray energies, luminosities and intrinsic times have been computed in the rest-frame 0.3-30 keV energy band. All the
other X-ray quantities refer to the 0.3-10 (observer frame) band.}										
\label{tab_histo}
\begin{tabular}{l|cccccccc}
\hline
\hline
 & $\#$ & $m$ & $M$ & $SD$ & $SK$ & $\mu$ & $\sigma$ & $N$ \\
\hline
$Log(S_{\gamma})$  & $ 386 $ & $ 0.17 $ & $ 0.18 $ & $ 0.61 $ & $ -0.07 $ & $ 0.17 \pm 0.05 $ & $ 0.60 \pm 0.04 $ & $ 105.1 \pm 7.7 $ \\
$Log(T_{90})$  & $ 334 $ & $ 1.58 $ & $ 1.73 $ & $ 0.65 $ & $ -1.18 $ & $ 1.67 \pm 0.07 $ & $ 0.62 \pm 0.06 $ & $ 93.9 \pm 9.4 $ \\
$Log(E_{\gamma}^{15-150})$  & $ 151 $ & $ 2.06 $ & $ 2.37 $ & $ 0.93 $ & $ -1.12 $ & $ 2.26 \pm 0.14 $ & $ 1.02 \pm 0.1 $ & $ 91.6 \pm 9. $ \\
$Log(E_{\rm{\gamma,iso}})$  & $ 78 $ & $ 2.88 $ & $ 3.01 $ & $ 0.91 $ & $ -0.89 $ & $ 3.01 \pm 0.14 $ & $ 0.85 \pm 0.12 $ & $ 38.9 \pm 5.7 $ \\
$Log(E_{\rm{pk}})$  & $ 78 $ & $ 2.64 $ & $ 2.71 $ & $ 0.52 $ & $ -0.76 $ & $ 2.56 \pm 0.1 $ & $ 0.47 \pm 0.069 $ & $ 23.6 \pm 4.7 $ \\
$Log(L_{\rm{pk,iso}})$  & $ 85 $ & $ 2.43 $ & $ 2.51 $ & $ 1.00 $ & $ -1.90 $ & $ 2.51 \pm 0.19 $ & $ 0.91 \pm 0.12 $ & $ 61.0 \pm 9.3 $ \\
$Log(T^{\rm{RF}}_{90})$  & $ 138 $ & $ 1.18 $ & $ 1.33 $ & $ 0.59 $ & $ -0.65 $ & $ 1.26 \pm 0.11 $ & $ 0.67 \pm 0.09 $ & $ 53.8 \pm 7.4 $ \\
$Log(\rm{NH_{HG}})$  & $ 161 $ & $21.6$ & $21.8$ & $ 1.21 $ & $ -3.54 $ & $21.9 \pm 0.1 $ & $ 0.62 \pm 0.09 $ & $ 42.5 \pm 7.3 $ \\
$Log(S_{\rm{X}})$ &  $316$  &    $-0.38$&    $-0.42$  &  $0.62$ & $ 0.23$&   $ -0.46 \pm0.05$ &   $0.57\pm0.04$ &  $ 68.6   \pm 5.3$ \\
$Log(S_{\rm{X}}^{\rm{FL}})$  &$   115 $ &  $-0.81$&    $-0.76$ &   $0.82$  &  $-0.38$ &   $ -0.76  \pm 0.12$ &    $0.89 \pm 0.12 $ &   $51.3  \pm  6.0$ \\
$Log(S_{\rm{1,X}})$  & $   211$  &  $-0.90$&    $-0.92$ &   $ 0.80$&$ 0.12$&   $-0.89 \pm 0.10$&    $0.88\pm0.06$ &    $97.2 \pm 7.9$\\
$Log(S_{\rm{2,X}})$  &    $316$&    $-0.58$  &  $-0.61$  &  $0.63$ & $0.14$ &    $-0.57\pm0.07$ &   $ 0.70\pm0.06$&    $102.2\pm8.5$\\
$Log(S_{\rm{1,X}}^{\rm{FL}})$  &    $62$ &   $-0.75$ &    $-0.69$ &   $0.73$&   $-0.38$   & - & - & -   \\
$Log(S_{\rm{2,X}}^{\rm{FL}})$  & $ 71 $ & $ -1.08 $ & $ -0.98 $ & $ 0.90 $ & $ 0.03 $ & $ -1.08 \pm 0.17 $ & $ 0.95 \pm 0.16 $ & $ 36.7 \pm 5.6 $ \\
$Log(E_{\rm{X,iso}})$  & $ 126 $ & $ 1.67 $ & $ 1.84 $ & $ 0.81 $ & $ -0.67 $ & $ 1.82 \pm 0.08 $ & $ 0.88 \pm 0.08 $ & $ 31.1 \pm 2.5 $ \\
$Log(E_{\rm{X}}^{\rm{FL}})$  & $ 59 $ & $ 1.25 $ & $ 1.40 $ & $ 0.97 $ & $ -0.64 $ & - & - & -   \\
$Log(E_{\rm{1,X}})$  & $ 86 $ & $ 1.00 $ & $ 1.04 $ & $ 0.92 $ & $ -0.28 $ & $ 1.10 \pm 0.11 $ & $ 0.94 \pm 0.08 $ & $ 40.7 \pm 3.6 $ \\
$Log(E_{\rm{2,X}})$  & $ 126 $ & $ 1.45 $ & $ 1.63 $ & $ 0.92 $ & $ -0.94 $ & $ 1.63 \pm 0.11 $ & $ 0.82 \pm 0.10 $ & $ 63.2 \pm 7.0 $ \\
$Log(E_{\rm{1,X}}^{\rm{FL}})$  & $ 35 $ & $ 1.13 $ & $ 1.38 $ & $ 1.00$ & $ -0.78 $ & - & - & - \\
$Log(E_{\rm{2,X}}^{\rm{FL}})$  & $ 38 $ & $ 1.04 $ & $ 1.14 $ & $ 0.98 $ & $ 0.01 $ & - & - & - \\
$Log(t_{\rm{i}})$  & $ 155 $ & $ 2.66 $ & $ 2.56 $ & $ 0.48 $ & $ 1.06 $ & - & - & -   \\
$Log(t_{\rm{f}})$  & $ 155 $ & $ 3.94 $ & $ 3.93 $ & $ 0.73 $ & $ 0.19 $ & $ 3.93 \pm 0.14 $ & $ 0.8 \pm 0.12 $ & $ 59.2 \pm 8.9 $ \\
$Log(F_{\rm{i}})$  & $ 155 $ & $ -4.23 $ & $ -4.26 $ & $ 0.83 $ & $ -0.14 $ & $ -4.18 \pm 0.11 $ & $ 0.89 \pm 0.12 $ & $ 61.1 \pm 6.6 $ \\
$Log(F_{\rm{f}})$  & $ 155 $ & $ -5.01 $ & $ -4.94 $ & $ 0.80 $ & $ -0.12 $ & $ -4.99 \pm 0.11 $ & $ 0.80 \pm 0.08 $ & $ 63.4 \pm 6.9 $ \\
$Log(t_{\rm{i}}^{\rm{RF}})$  & $ 62 $ & $ 2.13 $ & $ 2.03 $ & $ 0.62 $ & $ 1.27 $ & - & - & -   \\
$Log(t_{\rm{f}}^{\rm{RF}})$  & $ 62 $ & $ 3.58 $ & $ 3.48 $ & $ 0.74 $ & $ 0.29 $ & $ 3.53 \pm 0.11 $ & $ 0.75 \pm 0.09 $ & $ 23.0 \pm 2.9 $ \\
$Log(L_{\rm{i}})$  & $ 62 $ & $ 0.54 $ & $ 0.73 $ & $ 1.25 $ & $ -1.07 $ & $ 0.79 \pm 0.19 $ & $ 1.26 \pm 0.16 $ & $ 46.6 \pm 5.5 $ \\
$Log(L_{\rm{f}})$  & $ 62 $ & $ -0.47 $ & $ -0.19 $ & $ 1.19 $ & $ -1.19 $ & $ 0.04 \pm 0.33 $ & $ 1.41\pm 0.29 $ & $ 38.2 \pm 6.8 $ \\
\hline
$\alpha_{\rm{st}}$  & $ 213 $ & $ 3.96 $ & $ 3.56 $ & $ 2.34 $ & $ 3.95 $ & $ 3.22 \pm 0.51 $ & $ 2.34 \pm 0.35 $ & $ 238.0 \pm 35.0 $ \\
$\alpha_{\rm{sh}}$  & $ 155 $ & $ -0.16 $ & $ 0.18 $ & $ 1.23 $ & $ -4.06 $ & $ 0.27 \pm 0.14 $ & $ 0.52 \pm 0.12 $ & $ 61.0 \pm 13.0 $ \\
$\alpha_{\rm{n}}$  & $ 204 $ & $ 1.59 $ & $ 1.38 $ & $ 1.04 $ & $ 8.30 $ & $ 1.34 \pm 0.13 $ & $ 0.49 \pm 0.11 $ & $ 93.0 \pm 19.0 $ \\
\hline
\end{tabular}
\end{table*}

\begin{figure}
\includegraphics[width=1. \hsize,clip]{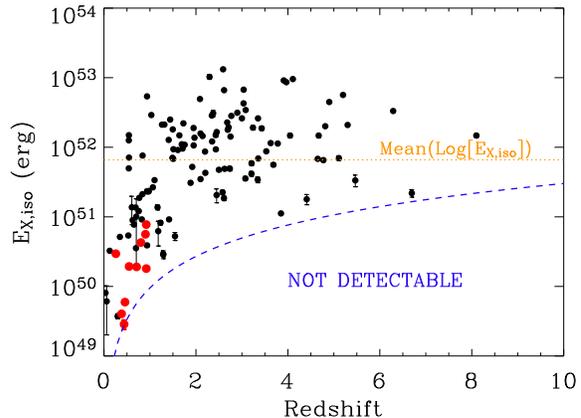}
\caption{0.3-30 keV (rest-frame) X-ray energy as a function of redshift. Black (red) points: long (short) GRBs.
Blue dashed line: empirically derived detectability threshold.
For $z>2$ we are not sensitive to GRBs with $E_{\rm{X,iso}}\lesssim 10^{51}$ erg. 
For $z>1$ there is no evidence for an evolution of the upper bound of $E_{\rm{X,iso}}$
with redshift. }
 \label{Fig:Exz}
\end{figure}


\subsection{Energetics of long and short GRBs}
\label{SubSubSec:Energetics}

Table \ref{tab_histo} reports the analysis of the 
parameter distributions derived from the LC fitting of Sect. \ref{Sec:dataanalysis}.
A complete list of symbols can be found in Appendix
\ref{Appendix:par}.  
The observed $E_{\rm{X,iso}}$ distribution peaks at $\sim7\times 10^{51}$ erg, 
typically representing $\sim7\%$ of the $1-10^4$ keV (rest-frame) $E_{\rm{\gamma,iso}}$. 
Figure \ref{Fig:Exz} shows that we are not sensitive to the population of bursts with 
$E_{\rm{X,iso}}<10^{51}$ erg for $z>2$ (so that the low energy tail of the $E_{\rm{X,iso}}$ distribution
is currently under-sampled). This is likely a non-detectability zone, consequence 
of the $E_{\rm{X,iso}}\propto E_{\rm{\gamma,iso}}^{0.8}$ of Sec. \ref{SubSec:parcor}.
For $z>1$ there is no evidence for an evolution of the 
upper bound of $E_{\rm{X,iso}}$ with redshift, which may suggest that $\sim10^{53}\rm{erg}$ 
is a physical boundary to the $E_{\rm{X,iso}}$ distribution 
(the record holder is GRB\,080721 with $E_{\rm{X,iso}}\sim10^{53}$ erg). In this respect we note that 
maximum budget $E_{max}\sim10^{52}$ erg\footnote{It is not a given that GRB\,080721 violates
this limit: $E_{\rm{X,iso}}$ represents the \emph{isotropic} equivalent X-ray energy, an 
overestimate to the true value if the emission -as we believe- is beamed.} is predicted by magnetar models \citep{Usov92}.
The same pattern is followed by the flare energy $E_{\rm{X}}^{\rm{FL}}$: for $z>2$
we are not sensitive to $E_{\rm{X}}^{\rm{FL}}<10^{50}$ erg.

Observations suggest that the GRB X-ray LCs consist of two distinct phases
(see e.g. \citealt{Willingale07}): 
a first steep decay phase tightly connected to the prompt $\gamma$-ray emission 
(\citealt{Tagliaferri05}; \citealt{Goad06}); and a second phase characterized by a flattening of the
LC (with limited evidence for spectral evolution, see e.g. \citealt{Liang07}) 
followed by a ``normal decay" phase.  
Type IIa GRBs (Fig. \ref{Fig:lctype}) clearly show the presence of both components, 
with energy $E_{\rm{1,X}}=E_1$ and $E_{\rm{2,X}}=E_2+E_3$, 
respectively; for type Ia LCs, the lack of spectral evolution and the typically
mild slope $\alpha_1$ resembling $\alpha_2$ of type IIa  lead us to identify 
$E_{\rm{2,X}}=E_1+E_2$; type Ib GRBs show strong spectral evolution during the first LC segment: this
together with the transition to a milder decay at $t_{b,1}$ lead us to define $E_{\rm{1,X}}=E_1$ and 
$E_{\rm{2,X}}=E_2$;
in type IIb LCs the spectral and temporal properties of the first segment (with slope $\alpha_1$)  
strongly suggest that \emph{Swift}-XRT caught the end of the prompt emission in the X-rays: we therefore
define $E_{\rm{1,X}}=E_1+E_2$ and $E_{\rm{2,X}}=E_3$; the same is true for type III GRBs: in this case we define
$E_{\rm{1,X}}=E_1+E_2$, $E_{2,X}=E_3+E_4$. 
The two phases release comparable energy (see Table \ref{tab_histo}),
with $E_{\rm{1,X}}$ and $E_{\rm{2,X}}$ peaking at $\sim1.1\times10^{51}$ erg and $\sim4\times10^{51}$ erg, respectively.

In each distribution, 
short GRBs populate the low energy tail: $E_{\rm{X,iso}}^{short}\sim 10^{50}$ erg,
which is approximately 2 orders of magnitude \emph{less} than a typical long GRB.
Figure \ref{Fig:Exz} also shows that short GRBs are less energetic than long GRBs
in the \emph{same} redshift bin.
A systematic difference between the -still poorly constrained- jet opening angles of long and 
short GRBs, with short GRBs being less collimated than long GRBs, could in principle mitigate this energy gap.
If we compare the energy released during the two phases separately (i.e. early steep decline vs.
plateau plus subsequent decay), we find an indication that short GRBs
are more energetically deficient during the second phase then in the first phase,
i.e. $E_{\rm{X,2}}^{short}/E_{X,2}^{long}\sim0.014$ and 
$E_{\rm{X,1}}^{short}/E_{X,1}^{long}\sim0.054$.
This argues against a beaming related explanation, since the jet opening angles 
of long and short GRBs are expected to be more similar at late than at early times.
Short GRB light-curves decay faster than long GRBs
in the X-rays, typically resulting in shorter observations ($t_{end}\sim10^4$ s vs.  $t_{end}\sim10^5$ rest-frame): 
however, using the average $L\propto t^{-1.3}$ scaling above, we find 
$E_{\rm{X,2}}^{short}(t<10^4\,\rm{s})\sim E_{\rm{X,2}}^{short}(t<10^5\,\rm{s})$. 
Thus the relatively lower measured energy of the later LC phase in short GRBs 
compared to long GRBs is not due to the shorter observations.
\subsection{Intrinsic neutral hydrogen absorption in long and short GRBs}
\label{SubSubSec:nh}

\begin{figure}
\includegraphics[scale=0.8,width=1 \hsize,clip]{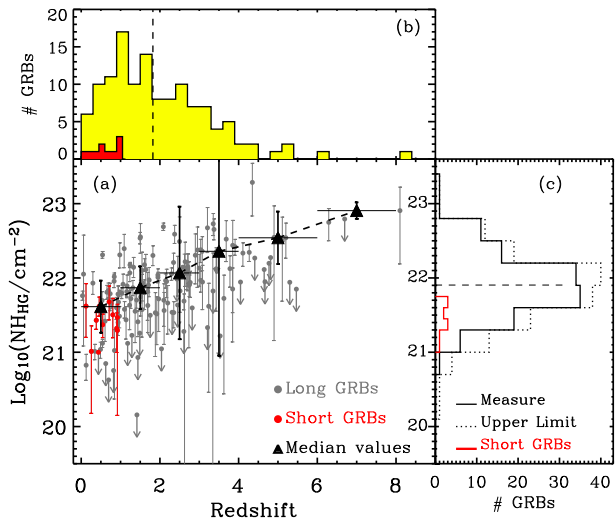}
\caption{\emph{Left panel}: 
\emph{(a)}: Intrinsic neutral hydrogen absorption vs. redshift for long and short GRBs (grey and red dots, respectively).
90\% upper limits are marked with arrows. Median $\rm{NH}_{\rm{HG}}$ values in different redshift bins are indicated with
filled triangles: for each bin, the error bars span the $1\sigma$ $\rm{NH}_{\rm{HG}}$ dispersion. \emph{(b)}:
redshift distribution of the sample of long (yellow) and short (red) GRBs. \emph{(c)}: Intrinsic neutral hydrogen distribution
for long (black line) and short (red line) GRBs. The dashed histogram includes upper limits.  In panels \emph{(b)}  and \emph{(c)} 
a dashed line indicates the median value for the entire distributions ($\langle z \rangle=1.82$, 
$\langle \rm{NH}_{\rm{HG}} \rangle=10^{21.8}\,\rm{cm^{-2}}$). } 
\label{Fig:NHvsz}
\end{figure}

The distribution of the intrinsic neutral hydrogen columns $\rm{NH_{HG}}$ is portrayed in 
Fig. \ref{Fig:NHvsz}, panel \emph{(c)}. The distribution of 
\emph{measured} $\rm{NH_{HG}}$ is found to have an average value\footnote{Solar abundances are used
to determine the best-fitting  $\rm{NH_{HG}}$.} 
$m=10^{21.6}\,\rm{cm^{-2}}$. When modeled
with a lognormal distribution, the best fitting mean and standard deviation are: $\mu=21.9\pm0.1$,
$\sigma=0.6\pm0.1$, in agreement with the estimates by \cite{Campana10,Campana11}  obtained on smaller samples.
However, differently from \cite{Campana10}: (i) our sample contains a larger number of GRBs for which
no evidence for intrinsic absorption was found (upper limits in Fig. \ref{Fig:NHvsz}); (ii) we find
evidence for a larger population of highly absorbed ($\rm{NH_{HG}}>10^{22}\rm{cm^{-2}}$) 
GRBs at low redshift ($z<2$). 

A trend for increasing $\rm{NH_{HG}}$ with redshift is apparent in Fig.  \ref{Fig:NHvsz}, panel \emph{(a)}:
however, our sensitivity to small amounts of intrinsic absorption decreases with increasing redshift
due to the fixed XRT band-pass, which explains the higher percentage of upper limits in the 4 to 6 redshift interval,
and is at least partially responsible for the observed trend. The sample is furthermore redshift selected, which
implies a bias against highly extinguished GRBs. The severity of this bias is possibly redshift
dependent, with a dependence which is difficult to quantify. 

Short GRBs map the low end of the $\rm{NH_{HG}}$ distribution, with an average absorption $
\rm{NH_{HG}^{short}}=10^{21.4}$ cm$^{-2}$ (mean of the logarithm of $\rm{NH_{HG}^{short}}$). 
Their properties are however \emph{consistent} with the intrinsic absorption of  \emph{long} GRBs in the same
redshift bin. A KS-test comparing the $\rm{NH_{HG}}$ distribution of long and short GRBs with 
$0<z<1$ reveals that there is \emph{no} evidence for long GRBs to show higher $\rm{NH_{HG}}$
when compared to short GRBs in the same redshift bin (KS probability of $34\%$ ).
A possibility is  that we missed the 
population of long GRBs with even higher $\rm{NH_{HG}}$ at low $z$ (see above). 
This would imply that GRBs with low optical extinction but high $\rm{NH_{HG}}$
are typical of the high-redshift universe, only (\citealt{Watson12}). 
We conclude that using the available data,
caution must be therefore used to interpret the long GRB $\rm{NH_{HG}}$ distribution as a proof of their 
association to star formation \citep{Campana10} unless this association is meant to be extended 
to short GRBs as well. 
\section{Parameter correlations}
\label{SubSec:parcor}

\begin{table*}
\centering
\caption{From left to right: X and Y parameters to be correlated 
(the best fitting law reads: $Log(Y)=q+mLog(X)$);
best-fitting parameters as obtained accounting for the sample variance \citep{Dagostini05}:
slope ($m$), normalization ($q$), intrinsic scatter ($\sigma$); errors are given at $95\%$ c.l.
The last six columns list the value of the Spearman rank $\rho$, Kendall coefficient $K$ and R-index $r$ statistics 
and relative chance probability $p$.
For each parameter couple, values reported in the first line refer to the entire sample, while in the second line 
we restrict our analysis to the long GRB class. X-ray fluences, fluxes and observer frame times are computed in 
the 0.3-10 keV (observer frame) energy band; luminosities, energies and rest-frame times are computed in the
0.3-30 keV (rest-frame) energy band.}				
\label{tab_corr}
\begin{tabular}{cc|ccccccccc}
\hline
\hline
$X$ & $Y$ & $m$ & $q$ & $\sigma$ & $\rho$ & $p(\rho)$ & $K$ & $p(K)$ & $r$ & $p(r)$ \\
\hline
$E_{\rm{X,iso}}$ & $E_{\rm{X}}^{\rm{FL}}$ & $1.07\pm0.03$ & $-4.1\pm72.2$ & $0.57\pm0.01$ & $0.70 $ & $3\times10^{-10}$ & $0.52$ & $2\times10^{-9}$ & $0.79$ & $<10^{-10}$ \\
& & $1.10\pm0.07$ & $-5.9\pm198$ & $0.58\pm0.01$ & $0.61$ & $3\times10^{-7}$ & $0.45$ & $6\times10^{-7}$ & $0.70$ & $10^{-9}$ \\
$E_{\rm{X,iso}}$ & $L_{\rm{f}}$ & $1.21\pm0.06$ & $-15.6\pm169$ & $0.85\pm0.01$ & $0.58$ & $6\times10^{-7}$ & $0.45$ & $2\times10^{-7}$ & $0.70$ & $3\times10^{-10}$ \\
& & $1.26\pm0.08$ & $-18.1\pm206$ & $0.85\pm0.01$ & $0.55$ & $3\times10^{-6}$ & $0.43$ & $8\times10^{-7}$ & $0.69$ & $10^{-9}$ \\
$E_{\rm{X,iso}}$ & $L_{\rm{i}}$ & $1.39\pm0.06$ & $-23.6\pm172$ & $0.83\pm0.01$ & $0.63$ & $3\times10^{-8}$ & $0.49$ & $2\times10^{-8}$ & $0.75$ & $<10^{-10}$ \\
& & $1.37\pm0.08$ & $-22.8\pm212$ & $0.84\pm0.02$ & $0.60$ & $3\times10^{-7}$ & $0.46$ & $2\times10^{-7}$ & $0.72$ & $10^{-10}$ \\
$t_{\rm{f}}^{\rm{RF}}$ & $L_{\rm{f}}$ & $-1.23\pm0.03$ & $51.9\pm0.46$ & $0.77\pm0.01$ & $-0.80$ & $<10^{-10}$ & $-0.60$ & $<10^{-10}$ & $-0.77$ & $<10^{-10}$ \\
& & $-1.24\pm0.03$ & $52.0\pm0.45$ & $0.73\pm0.01$ & $-0.82$ & $<10^{-10}$ & $-0.62$ & $<10^{-10}$ & $-0.78$ & $<10^{-10}$ \\
$L_{\rm{f}}$ & $E_{\rm{2,X}}$ & $0.52\pm0.01$ & $26.8\pm11.8$ & $0.47\pm0.00$ & $0.67$ & $2\times10^{-9}$ & $0.51$ & $5\times10^{-9}$ & $0.80$ & $<10^{-10}$ \\
& & $0.50\pm0.00$ & $27.6\pm10.8$ & $0.43\pm0.00$ & $0.65$ & $10^{-8}$ & $0.50$ & $2\times10^{-8}$ & $0.81$ & $<10^{-10}$ \\
$E_{\rm{2,X}}$ & $E_{\rm{1,X}}$ & $0.42\pm0.02$ & $29.1\pm43.2$ & $0.81\pm0.01$ & $0.42$ & $4\times10^{-5}$ & $0.29$ & $4\times10^{-5}$ & $0.45$ & $6\times10^{-6}$ \\
& & $-$ & $-$ & $-$ & $0.28$ & $6\times10^{-3}$ & $0.19$ & $0.06$ & $0.29$ & $6\times10^{-3}$ \\
$t_{\rm{f}}$ & $F_{\rm{f}}$ & $-0.79\pm0.01$ & $-7.80\pm0.09$ & $0.45\pm0.00$ & $-0.69$ & $<10^{-10}$ & $-0.50$ & $<10^{-10}$ & $-0.74$ & $<10^{-10}$ \\
& & $-0.79\pm0.01$ & $-7.78\pm0.09$ & $0.45\pm0.00$ & $-0.69$ & $<10^{-10}$ & $-0.50$ & $<10^{-10}$ & $-0.74$ & $<10^{-10}$ \\
$E_{\rm{\gamma,iso}}$ & $E_{\rm{X,iso}}$ & $0.79\pm0.01$ & $10.0\pm20.6$ & $0.39\pm0.00$ & $0.86$ & $<10^{-10}$ & $0.69$ & $<10^{-10}$ & $0.88$  & $<10^{-10}$\\
& & $0.67\pm0.01$ & $16.5\pm18.8$ & $0.29\pm0.00$ & $0.82$ & $<10^{-10}$ & $0.63$ & $<10^{-10}$ & $0.88$ & $<10^{-10}$ \\
$E_{\rm{\gamma,iso}}$ & $E_{\rm{X}}^{\rm{FL}}$ & $0.89\pm0.05$ & $3.85\pm148$ & $0.65\pm0.01$ & $0.64$ & $8\times10^{-5}$ & $0.48$ & $10^{-4}$ & $0.74$ & $3\times10^{-6}$ \\
& & $0.93\pm0.10$ & $1.83\pm287$ & $0.62\pm0.02$ & $0.56$ & $10^{-3}$ & $0.41$ & $10^{-3}$ & $0.67$ & $6\times10^{-5}$ \\
$E_{\rm{\gamma,iso}}$ & $E_{\rm{1,X}}$ & $0.67\pm0.03$ & $15.9\pm91.3$ & $0.81\pm0.02$ & $0.71$ & $2\times10^{-7}$ & $0.56$ & $2\times10^{-7}$ & $0.64$ & $5\times10^{-6}$ \\
& & $0.56\pm0.04$ & $21.6\pm126$ & $0.77\pm0.02$ & $0.62$ & $5\times10^{-5}$ & $0.48$ & $4\times10^{-5}$ & $0.54$ & $5\times10^{-4}$ \\
$E_{\rm{\gamma,iso}}$ & $E_{\rm{2,X}}$ & $0.92\pm0.01$ & $2.96\pm33.5$ & $0.51\pm0.01$ & $0.76$ & $<10^{-10}$ & $0.59$ & $<10^{-10}$ & $0.85$ & $<10^{-10}$ \\
& & $0.74\pm0.01$ & $12.6\pm35.1$ & $0.44\pm0.00$ & $0.67$ & $10^{-8}$ & $0.51$ & $2\times10^{-8}$ & $0.81$ & $<10^{-10}$ \\
$E_{\rm{\gamma,iso}}$ & $L_{\rm{f}}$ & $1.06\pm0.08$ & $-8.86\pm227$ & $1.03\pm0.04$ & $0.54$ & $9\times10^{-4}$ & $0.41$ & $7\times10^{-4}$ & $0.70$ & $8\times10^{-6}$ \\
& & $1.05\pm0.09$ & $-8.43\pm260$ & $1.06\pm0.04$ & $0.50$ & $3\times10^{-3}$ & $0.37$ & $3\times10^{-3}$ & $0.68$ & $2\times10^{-5}$ \\
$E_{\rm{pk}}$ & $E_{\rm{X,iso}}$ & - & - & - & - & - & -& - & - & - \\
& & $0.98\pm0.02$ & $49.5\pm0.15$ & $0.37\pm0.00$ & $0.63$ & $10^{-7}$ & $0.46$ & $5\times10^{-7}$ & $0.76$ & $<10^{-10}$ \\
$L_{\rm{pk}}$ & $E_{\rm{X,iso}}$ & - & - & - & - & - & - \\
& & $0.48\pm0.01$ & $27.0\pm16.4$ & $0.44\pm0.00$ & $0.58$ & $2\times10^{-7}$ & $0.42$ & $3\times10^{-7}$ & $0.74$ & $<10^{-10}$ \\
$L_{\rm{pk}}$ & $L_{\rm{f}}$ & - & - & - & - & - & - \\
& & $0.86\pm0.03$ & $2.35\pm87.6$ & $0.87\pm0.02$ & $0.50$ & $10^{-3}$ & $0.39$ & $5\times10^{-4}$ & $0.76$ & $7\times10^{-8}$ \\
$L_{\rm{pk}}$ & $E_{\rm{2,X}}$ & - & - & - & - & - & - \\
& & $0.60\pm0.01$ & $20.3\pm15.9$ & $0.43\pm0.00$ & $0.58$ & $10^{-7}$ & $0.42$ & $3\times10^{-7}$ & $0.82$ & $<10^{-10}$ \\
$S_{\rm{\gamma}}$ & $S_{\rm{X}}$ & $0.77\pm0.01$ & $-7.80\pm0.09$ & $0.45\pm0.00$ & $0.79$ & $<10^{-10}$ & $0.59$ & $<10^{-10}$ & $0.77$ & $<10^{-10}$ \\
& & $0.82\pm0.00$ & $-1.58\pm0.10$ & $0.37\pm0.00$ & $0.78$ & $<10^{-10}$ & $0.58$ & $<10^{-10}$ & $0.78$ & $<10^{-10}$ \\
\hline
$E_{\rm{\gamma,iso}}$ & $L_{X}^{\rm{11h}}$ & $0.71\pm0.01$&  $8.53\pm30.9$ & $0.55\pm0.01$  &  $0.66$  &  $3\times10^{-10}$  & $ 0.49$ &  $1\times10^{-9}$  &  $0.77$ &  $<10^{-10}$\\
& &                                                                                $0.54\pm0.01$& $17.5\pm29.6$&$0.45\pm0.00$&    $0.55$  &  $2\times10^{-6}$  &  $0.40$ &   $2\times10^{-6}$ &  $0.70$ & $<10^{-10}$   \\
$E_{\rm{\gamma,iso}}$ & $L_{X}^{\rm{10min}}$ &$0.93\pm 0.01$&$-1.17\pm0.01$&$0.45 \pm0.01$& $ 0.87$ &  $<10^{-10}$  &  $0.67$ &  $<10^{-10}$   &$ 0.88$ & $<10^{-10}$ \\
&  &                                                                                $0.78\pm 0.01$&$6.73\pm32.6$&$ 0.40\pm 0.00$&$0.82$ & $<10^{-10}$ &$0.63$ &  $<10^{-10}$  &    $0.84$ & $<10^{-10}$ \\
\hline
\end{tabular}
\end{table*}

Here we proceed to look for 2-parameter correlations involving both X-ray and $\gamma$-ray
properties. From a blind analysis we found 199 statistically significant correlations
(out of 946). We focus on the physically interesting, correlations. 
The significance of each correlation is estimated using the R-index $r$, the 
Spearman rank $\rho$ and Kendall coefficient $K$ (Table \ref{tab_corr}). 
Only correlations for which the chance probability associated with at least one of the 
test statistics is $<10^{-3}$ have been listed.
As a general note: 
\begin{itemize}
\item No significant correlation is found to involve the rest frame $T_{90}$, the intrinsic $\rm{{NH}_{HG}}$
or the LC temporal slopes;
\item We re-scaled the LC temporal breaks $t_{\rm{b}}$ by the $T_{90}$, adding new parameters
to our list: $y\equiv t_{\rm{b}}/T_{90}$. However, the use of re-scaled properties did not improve any of our correlations 
and are therefore not included in the following discussion. 
\end{itemize}

The correlation coefficients and the best-fitting power-law parameters of each 
correlation are listed in Table \ref{tab_corr}.
Our best-fitting procedure accounts for the sample variance \citep{Dagostini05}.

\begin{figure*}
\includegraphics[width=0.45 \hsize,clip]{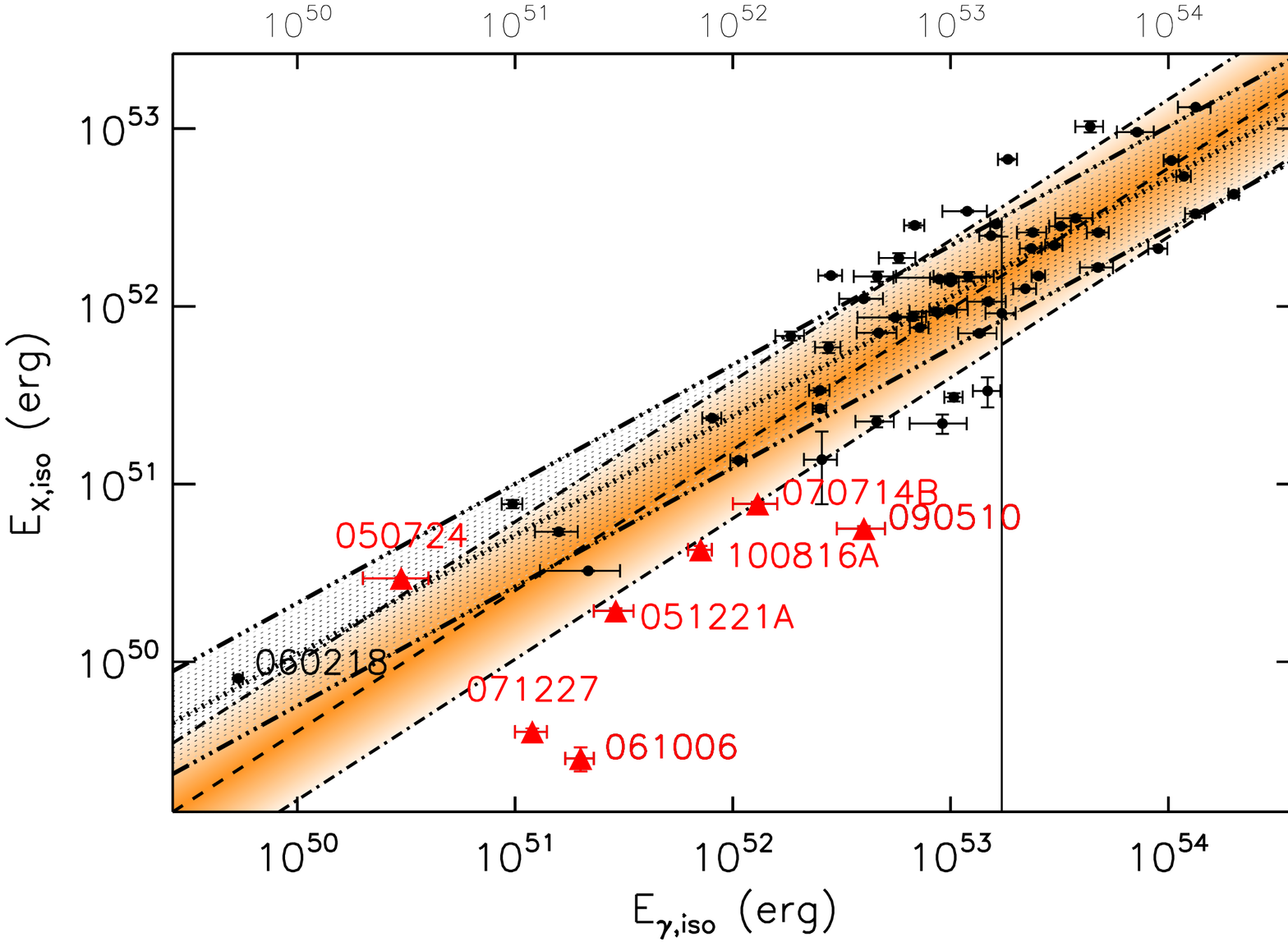}
\includegraphics[width=0.45 \hsize,clip]{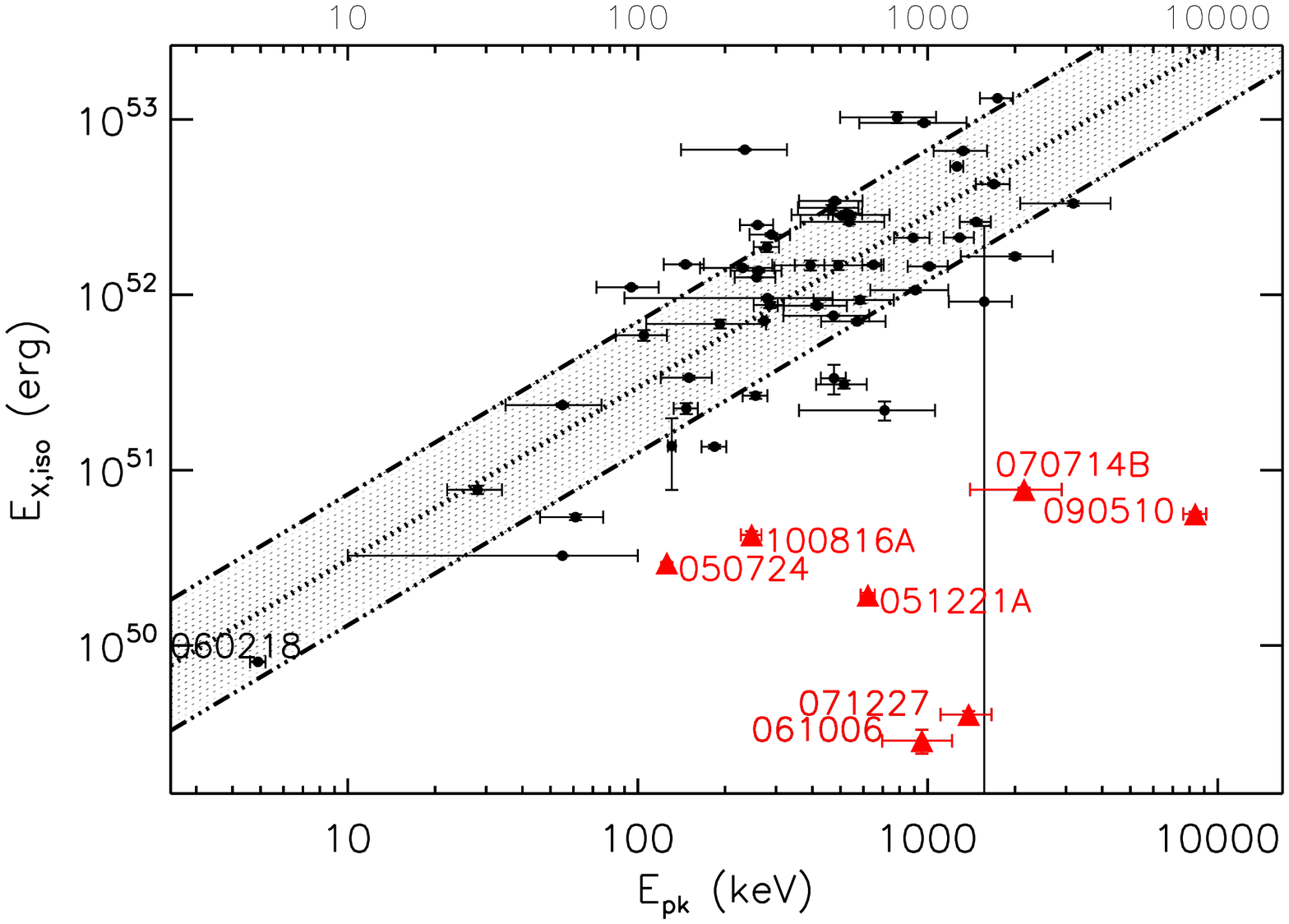}\\
  \caption{Correlations involving $E_{\gamma\rm{,iso}}$ ($1-10^4$ keV, rest-frame), $E_{\rm{X,iso}}$
  (0.3-30 keV, rest-frame) and the rest-frame prompt peak energy $E_{\rm{pk}}$.  Black
  dots (red triangles): long (short) GRBs. Dashed line: best-fitting power-law model for the entire short plus
  long GRBs sample. Dotted line: best-fitting model for the long GRB class, only. The coloured and hatched 
  areas mark the 68\% confidence region around the best fit. Short GRBs and outliers are 
  named.}
   \label{Fig:parcor2}
\end{figure*}

\subsubsection{The link between the X-ray and prompt $\gamma$-ray energy}
\label{SubSubSec:energycorr}

Fig. \ref{Fig:parcor2}, (left panel) shows that 
$E_{\rm{X,iso}}$ is directly linked to the isotropic energy released in $\gamma$-rays
during the prompt emission $E_{\rm{\gamma,iso}}$. A similar result was found by 
\cite{Willingale07} on a smaller sample of GRBs. Here we show for the first time how short GRBs compare
to long GRBs: notably, all short GRBs but GRB\,050724 are outliers of the long GRB relation, with
$E_{\rm{X,iso}}$ for short GRBs a factor $\sim 50$ below that for long GRBs and having
large dispersion (the $E_{\rm{X,iso}}$ 
distributions are almost distinct for long and short GRBs, as shown in Fig. \ref{Fig:Exz}).
A clear exception is GRB\,050724
(\citealt{Barthelmy05}; \citealt{Grupe06}) which had a bright and long-lived X-ray afterglow with a powerful late time 
re-brightening (\citealt{Bernardini11a}; \citealt{Campana06}; \citealt{Malesani07}). 
This difference may be understood in terms of a different radiative efficiency $\eta_{\gamma}$
(where $\eta_{\gamma}\equiv E_{\gamma}/(E_{\gamma}+ E_{K})$, being $E_{K}$ the outflow kinetic
energy) during the prompt emission
between short GRBs and XRFs (X-ray Flashes, i.e. GRBs with $E_{\rm{\gamma,iso}}\lesssim 10^{52}\,\rm{erg}$
in Fig. \ref{Fig:parcor2}): in this picture,  $\eta_{\gamma}^{short}> \eta_{\gamma}^{XRF}$. The two
populations are clearly distinct in terms of spectral peak energy during the prompt phase, with
$E_{\rm{pk}}^{short}>E_{\rm{pk}}^{XRFs}$ (Fig.  \ref{Fig:parcor2}). 
This may suggest that $\eta_{\gamma}$ anti-correlates with $E_{\rm{pk}}$: this is further
investigated in Sec. \ref{SubSubSec:3parcor}.

Short and long GRBs occupy different areas of the $E_{\rm{X,iso}}$ vs. $E_{\rm{pk}}$ plane (Fig. 
\ref{Fig:parcor2}, upper right panel) as well, demonstrating how the information from the X-ray 
LCs can be used to infer the GRB nature. Again, short GRBs fall below the long GRBs. 

\subsubsection{The X-ray plateau and the prompt $\gamma$-ray phase in long and short GRBs}
\label{SubSubSec:Lfenergy}

\begin{figure}
\includegraphics[width=1. \hsize,clip]{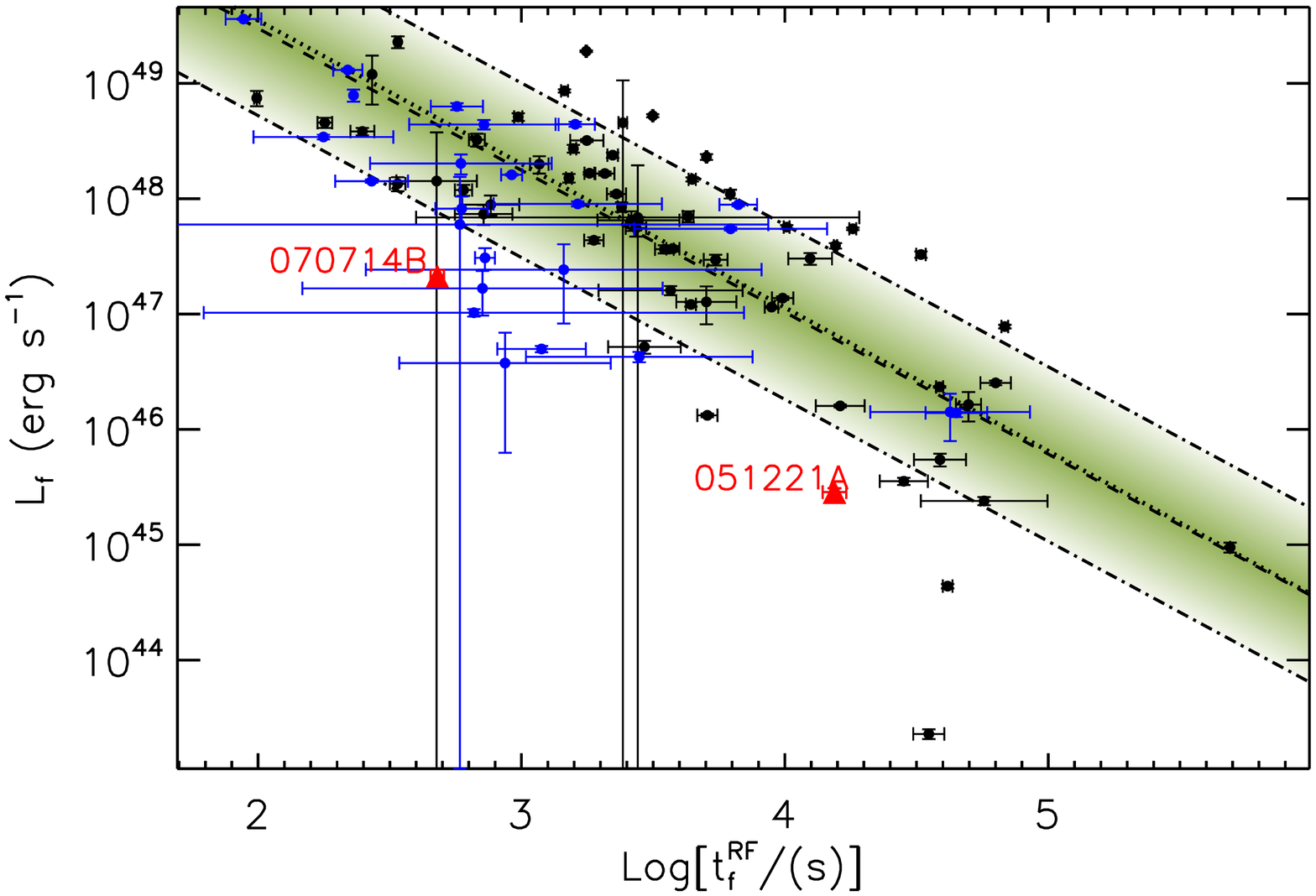}
  \caption{
  Luminosity at the end of the plateau phase, 0.3-30 keV (rest-frame) vs. end-time  of the plateau.
   Colour coding as in Fig.  \ref{Fig:parcor2}. Blue points are for type Ia LCs.}
   \label{Fig:parcor7}
\end{figure}

In the literature,  the shallow decay (or ``plateau") is associated with
an LC phase generally characterized by a mild slope (and absence of spectral
evolution in the X-rays; \citealt{Liang07}): this can be identified in type IIa and type III light-curves. In type IIa (III) GRBs, this phase
starts at $t_{\rm{i}} \equiv t_{\rm{b1}}$ ($t_{\rm{i}}\equiv t_{\rm{b2}}$) and ends at $t_{\rm{f}}\equiv t_{\rm{b2}}$ 
($t_{\rm{f}}\equiv t_{\rm{b3}}$), 
with temporal slope $\alpha_2$ ($\alpha_3$) and energy $E_2$ ($E_3$).
Short GRBs are under-represented in the class of GRBs showing clear evidence 
of plateaus in the X-rays. Only 2 short GRBs (out of 19 with C-like LCs\footnote{36 was the number of
short GRBs in our starting sample; only 19 of these have C-like LCs.}, $10\%$) 
possibly have plateaus:
GRB\,051221A ($T_{90}=1.4$ s) and GRB\,070714B ($T_{90}=3$ s, extended emission
not included).
The corresponding percentage for long GRBs is instead $\sim37$\%. 

The luminosity at the end of the plateau phase $L_{\rm{f}}$ is directly related to the total energy released in the 
second LC phase $E_{\rm{2,X}}$ (Table \ref{tab_corr}): $E_{\rm{2,X}}\propto L_{\rm{f}}^{0.52}$. 
It is interesting to note that of the two short GRBs, 070714B is a clear outlier, while 051221A is only barely 
consistent with the correlation. The peculiar GRB\,060218 also shows a lower than expected $E_{\rm{2,X}}$. 
\cite{Dainotti08} first reported a correlation between $L_{\rm{f}}$ and $t_{\rm{f}}^{\rm{RF}}$ for 
long GRBs. Here we confirm the correlation (with best-fitting $L_{\rm{f}}\propto (t_{\rm{f}}^{\rm{RF}})^{-1.2}$ 
Fig. \ref{Fig:parcor7}) and show that the two short GRBs with clear evidence of plateau are \emph{not}
consistent with the same scaling.

\subsubsection{The link between the X-ray luminosity and the prompt $\gamma$-ray energy release}
\label{SubSec:LEiso}

\begin{figure*}
\includegraphics[scale=0.38]{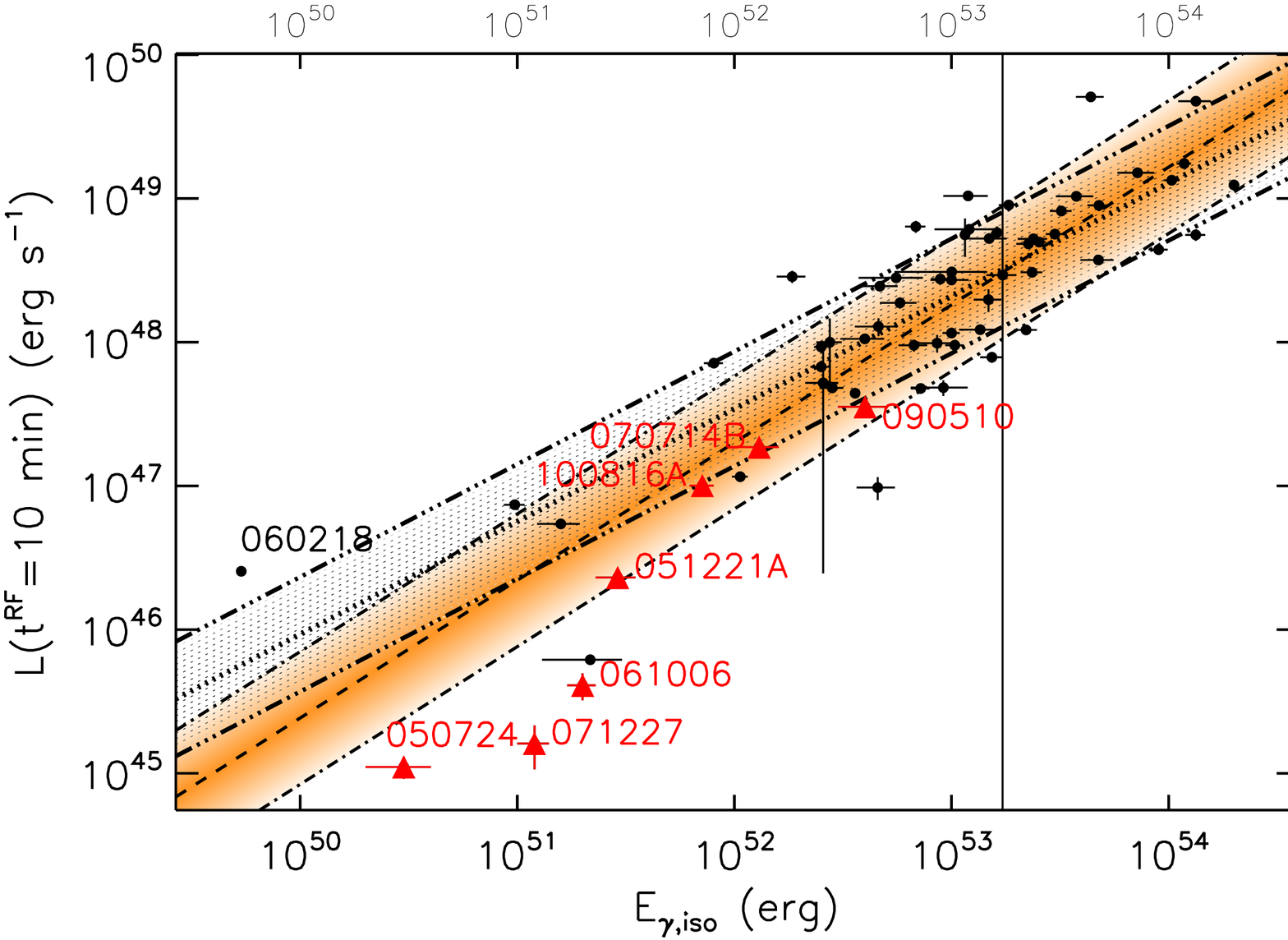}
\includegraphics[scale=0.38]{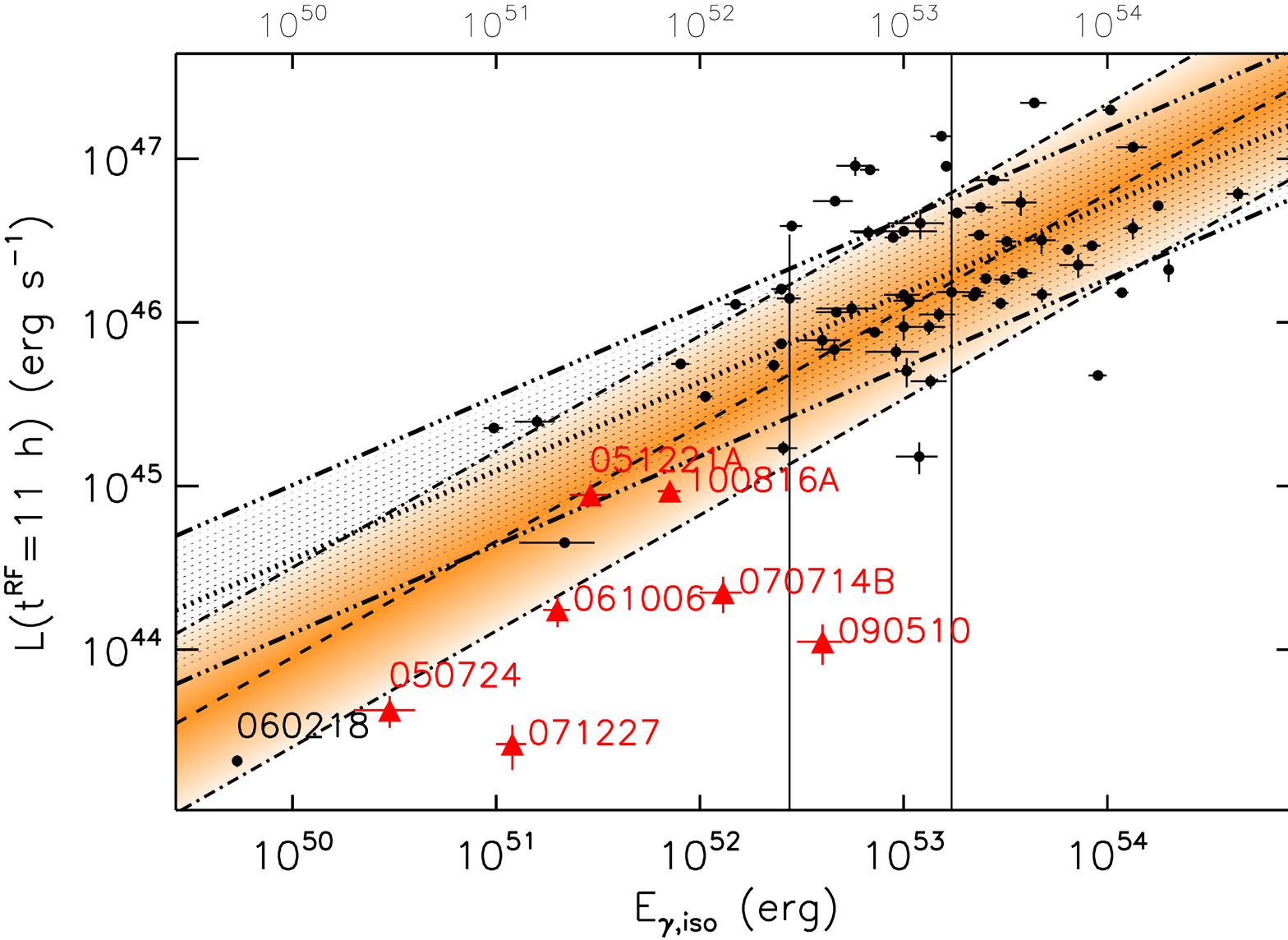}
  \caption{0.3-30 keV (rest-frame) X-ray LC luminosity measured at 10 minutes (left panel) and 11hr (right panel)
  rest-frame vs. $E_{\rm{\gamma,iso}}$. Colour coding as in Fig.  \ref{Fig:parcor2}. The scatter of the
  correlation increases with time.}
\label{Fig:LEiso}
\end{figure*}

The X-ray luminosity of the LC, $L_{\rm{X}}(t^{\rm{RF}})$, correlates with the $\gamma$-ray 
energy released during the prompt emission for \emph{any} $t^{\rm{RF}}$ between $100$ s and $10^5$ s. 
Here we arbitrarily select two rest-frame times (Fig.  \ref{Fig:LEiso})
as an example. We find that the scatter of the correlation evolves with time,
with the $L_{\rm{X}}(t^{\rm{RF}})$ vs. $E_{\rm{\gamma,iso}}$ being tighter at early times (see Fig.  \ref{Fig:LEiso}). 
For this plot we require the GRBs to have been \emph{observed} at those rest-frame
times  but relax the LC completeness requirement. No extrapolation of the observed LC is performed.

At early times the LC luminosity tracks $E_{\rm{\gamma,iso}}$ with limited dispersion around the best-fitting
model $L_{\rm{X}}^{\rm{10min}}\propto E_{\rm{\gamma,iso}}^{0.9}$. Short GRBs tend to lie below the 
best-fitting law of the long GRB class. When compared to the same relation at much later times (11 hours) we find that:
(i) the relation is now more scattered, suggesting that the X-ray LCs are more directly linked to the prompt $\gamma$-ray
phase at early than at late times; (ii) while the relation is  highly scattered, we note that all short GRBs of our sample 
lie below the long GRB relation: this is consistent with the 
steeper decay of the average short GRBs LC when compared to the long GRBs LC found in Sec \ref{SubSec:aftdis}.
Our analysis therefore does not confirm the previous results from \cite{Nysewander09}, who found that 
short and long GRBs are consistent with the \emph{same} $L_{\rm{X}}$ vs. $E_{\rm{\gamma,iso}}$ scaling
(note however that their $E_{\rm{\gamma,iso}}$ is computed in a much narrower energy band).
\subsubsection{Observational biasses:  temporal extrapolation}
\label{SubSec:ObsBias}

\begin{figure}
\includegraphics[scale=0.4]{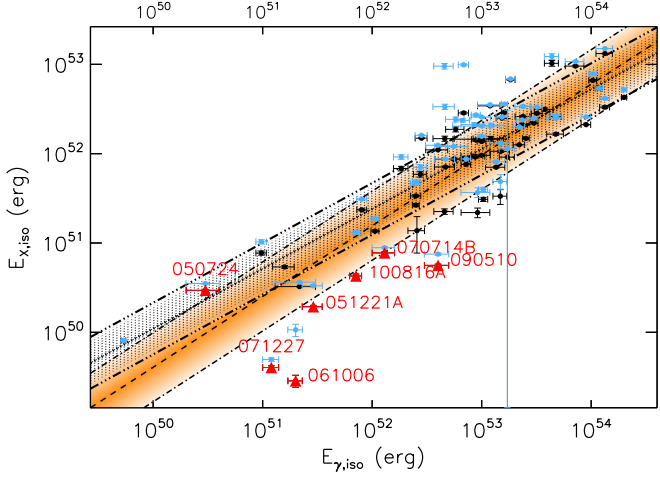}
  \caption{Impact of the temporal extrapolation of the observed LCs on the
  $E_{\rm{X,iso}}$ vs. $E_{\gamma,iso}$ correlation. Colour coding as in Fig.  \ref{Fig:parcor2}.
  Light blue dots: $E_{\rm{X,iso}}$ values have been computed integrating the luminosity
  over a  common rest-frame interval of time.}
\label{Fig:kcorr}
\end{figure}
 
The \emph{Swift} re-pointing time $t_{\rm{rep}}$ and end time of the observations $t_{\rm{end}}$ 
 vary from GRB to GRB. Since $E_{\rm{X,iso}}$ is obtained by integrating  the luminosity of each LC between
$t_{\rm{rep}}$ and $t_{\rm{end}}$, one may wonder what is the effect of using different integration
times for different GRBs. This is quantified as follows.
To estimate the amount of energy lost at the end of the observations, 
we extrapolated the best-fitting profile of each GRB up to $10^7\,\rm{s}$
(rest frame) and integrated the LC luminosity up until that time.  
Since GRBs may experience a jet break at late times \citep{Racusin09}, this computation may lead to an 
overestimate of the real energy lost. 
The amount of energy possibly lost at the beginning of the observations\footnote{Note 
that the sample of C-like GRBs we use to look for correlations was 
pre-selected requiring an \emph{observed} time of re-pointing $t_{obs}<300\,\rm{s}$
to minimize this effect.} is estimated by conservatively extrapolating backwards in time the best-fitting profile 
to the minimum rest-frame \emph{Swift} re-pointing time of our sample, which is 12.5 s. 
For GRBs with $T_{90}^{\rm{RF}}>12.5\,\rm{s}$, we adopt $T_{90}^{\rm{RF}}$
as the starting time for the integration to avoid extrapolating the luminosity to unrealistic values.
This approach leads to an \emph{overestimate} of the amount of energy lost before $t_{\rm{rep}}$
for the large majority of GRBs, as can be seen comparing the extrapolated temporal profile we are
adopting here to the \emph{Swift}-BAT emission at the same rest-frame time (see e.g. the
Swift Burst Analyser BAT plus XRT LCs of GRB\,050724, \citealt{Evans11}).
The corrected $E_{\rm{X,iso}}$ is shown in Fig.  \ref{Fig:kcorr} (light blue points).  Larger corrections
(up to a factor $\sim9$  for  GRB\,090510) are found to be applied to short GRBs.
In spite of the very conservative approach we find that short GRBs are still either barely compatible or not consistent
with the long GRB relation (as before), while the long GRBs relation is almost unaffected
by this correction.
 
We therefore conclude that in a logarithmic space the different rest-frame integration
time used does not create or destroy correlations.
The $E_{\rm{X,iso}}$ vs. $E_{\rm{\gamma,iso}}$ correlation 
has been used here as an example: this result  applies to all the  
relations presented in this paper.
\subsection{Multiparameter correlations}
\label{SubSec:multiparcor}

\begin{figure*}
\includegraphics[scale=0.4]{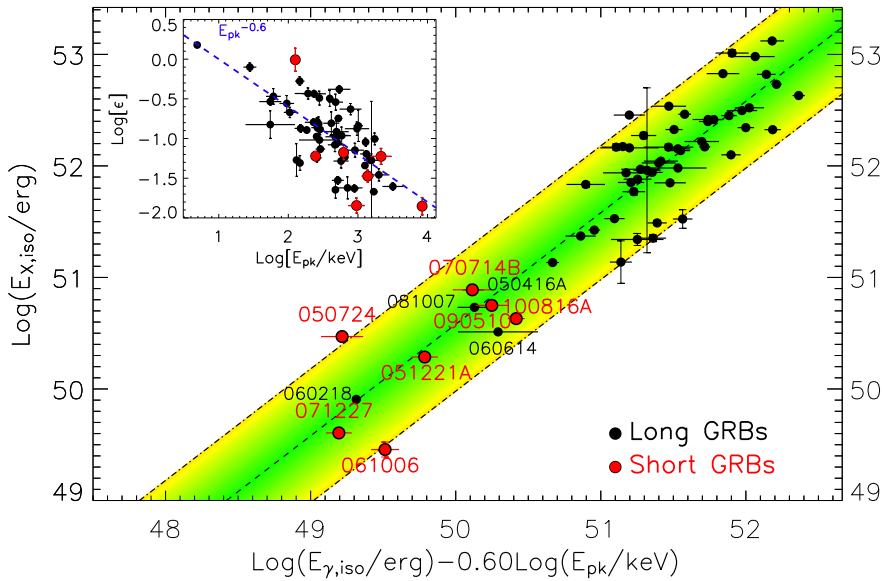}
  \caption{Three parameter correlation involving $E_{\rm{X,iso}}$, $E_{\rm{\gamma,iso}}$ and $E_{\rm{pk}}$.  Dashed line: best-fitting relation (Eq. \ref{Eq:3parcor}); dot-dashed lines mark the 95\% confidence area around the best-fitting law. Notably, long and short GRBs share the same scaling, with short and sub-energetic GRBs (like GRB\,060218) occupying the  same area of the plot. \emph{Inset:} evolution of the efficiency parameter $\epsilon\equiv E_{\rm{X,iso}}/E_{\rm{\gamma,iso}}$ as a function of the spectral peak energy of the prompt emission $E_{\rm{pk}}$. A reference $\epsilon \propto E_{\rm{pk}}^{-0.6}$ scaling has been marked with a blue dashed line.}
\label{Fig:3parcor}
\end{figure*}


We look here for correlations involving more than 2 parameters (either from the
X-rays or from the $\gamma$-rays). We first discuss the results from a Principal 
Component Analysis (PCA, Sect.  \ref{SubSubSec:PCA}) and then show the existence 
of a tight 3-parameter correlation directly linking $E_{\rm{X,iso}}$, $E_{\rm{\gamma,iso}}$ 
and $E_{\rm{pk}}$ both in long and short GRBs (Sect. \ref{SubSubSec:3parcor}).
\subsubsection{Principal Component Analysis}
\label{SubSubSec:PCA}

The Principal Component Analysis (PCA) is a statistical technique designed to find patterns 
in data: it uses orthogonal transformations to convert a set of possibly correlated variables 
into linearly uncorrelated (orthogonal) variables.
Given a set  of N events (GRBs in our case) described by M parameters, 
the PCA consists of the diagonalization of the covariance matrix: 
the eigenvenctors found are called principal components (PCs), while the eigenvalues 
consist of the variance associated  with each PC \citep[see][for details]{Jolliffe02}.
We performed a standardized PCA as recommended when the parameters have 
widely different variances: each parameter $P$ is replaced by  
$\hat{P}=(Log[P]-\overline{Log[P]})/\sigma_{Log[P]}$.
In this case the matrix to be diagonalized is not a covariance, but a correlation matrix. 
Calculations were performed using the statistical package 
\textbf{R}\footnote{http://www.r-project.org/}.

In Sect. \ref{SubSec:parcor} we showed that $E_{\rm{X,iso}}$ is the X-ray parameter that
still keeps information from the prompt $\gamma$-ray energy release. We now investigate its relation 
to other prompt parameters, specifically $E_{\rm{pk}}$, $L_{\rm{pk}}$, $T_{90}^{\rm{RF}}$ 
(and $E_{\rm{\gamma,iso}}$), using the PCA. This set of parameters is measured simultaneously 
in $44$ GRBs. Table~\ref{tab_pca_prompt+aft} reports the three most significant PCs 
($86\%$ of the total variance) projected upon the original $5$ variables. 
Each variable roughly contributes with comparable weight to the first PC; the second PC is instead 
dominated by  $\hat{T}_{90}^{\rm{RF}}$.
The third PC relates $\hat{E}_{\rm{X,iso}}$ with $\hat{E}_{\rm{pk}}$.  This result
suggests that, while $E_{\rm{pk}}$, $L_{\rm{pk}}$, $E_{\rm{X,iso}}$ and 
$E_{\rm{\gamma,iso}}$ are in some way physically related to one another
(see Sect. \ref{SubSubSec:3parcor}),  the duration of the $\gamma$-ray 
energy release represents an \emph{additional} degree of freedom 
to the system.

\begin{table}
\centering
\caption{The three most significant PCs ($85\%$ of the total variance) 
projected upon $\hat{E}_{\rm{\gamma,iso}}$, $\hat{E}_{\rm{pk}}$, 
$\hat{L}_{\rm{pk}}$, $\hat{T}_{90}^{\rm{RF}}$ and $\hat{E}_{\rm{X,iso}}$.}
\label{tab_pca_prompt+aft}
\begin{tabular}{l|ccc}
\hline
\hline
 & PC1 & PC2 & PC3\\
 & $40\%$ & $66\%$ & $85\%$\\
\hline
$\hat{E}_{\gamma,iso}$ & $-0.561$ & $0.141$  & $0.171$\\
$\hat{E}_{pk}$ & $-0.448$ & $-0.300$  & $-0.630$\\
$\hat{L}_{pk}$ & $-0.502$ & $-0.389$ & $-$\\
$\hat{T}_{90}^{RF}$ & $-0.121$ & $0.794$ & $-0.512$\\
$\hat{E}_{X,iso}$ & $-0.466$ & $0.331$ & $0.588$ \\
\hline
\end{tabular}
\end{table}
\subsubsection{A GRB universal scaling: $E_{\rm{X,iso}}$, $E_{\rm{\gamma,iso}}$ and $E_{\rm{pk}}$}
\label{SubSubSec:3parcor}

We look for a 3-parameter correlation
involving $E_{\rm{X,iso}}$, $E_{\rm{\gamma,iso}}$ and $E_{\rm{pk}}$. The three variables
are found to be correlated (see Fig. \ref{Fig:3parcor}) with the following  best fitting law
(obtained following the method by \citealt{Dagostini05}):

\begin{equation}
\label{Eq:3parcor}
E_{\rm{X,iso}}=10^{(0.58\pm0.25)}\Big[ \frac{E_{\rm{\gamma,iso}}^{(1.00\pm0.06)}}{E_{\rm{pk}}^{(0.60\pm0.10)}} \Big]\\
\end{equation}
where $E_{\rm{X,iso}}$, $E_{\rm{\gamma,iso}}$ and $E_{\rm{pk}}$  are in units of ergs
and keV, respectively. 
The intrinsic scatter is $\sigma_{\rm{E_{x,iso}}}=0.30\pm0.03$ ($1\,\sigma$). We note that:
\begin{itemize}
\item This relation expands on the well known $E_{\rm{pk}}$- $E_{\rm{\gamma,iso}}$
(\citealt{Amati08} and references therein) relation with the introduction of 
a third parameter ($E_{\rm{X,iso}}$).
\item It combines information from the prompt \emph{and} from the
X-ray energy release which \emph{follows} the prompt. 
While short GRBs are clear outliers of the $E_{\rm{pk}}$- $E_{\rm{\gamma,iso}}$
relation, they perfectly fit into the $E_{\rm{X,iso}}$-$E_{\rm{\gamma,iso}}$-$E_{\rm{pk}}$ relation:
the importance of the 3-parameter relation is that it combines short and long GRBs on a common
scaling. As a result, considering the \emph{entire} short plus long GRB sample, the scatter is \emph{reduced} 
by the introduction of the third variable (the intrinsic scatter of the Amati relation
of our sample of long and short GRBs is $\sigma_{\rm{E_{pk}}}=0.37$ to be compared to the intrinsic scatter of 
the 3-paramter relation on $E_{\rm{pk}}$ which is $\sigma_{\rm{E_{pk}}}=0.29$). 
Restricting our analysis to long GRBs, we find $\sigma_{\rm{E_{pk}}}=0.17$ both 
for the $E_{\rm{pk}}$- $E_{\rm{\gamma,iso}}$ and for the 3-parameter correlation.
\item Short GRBs (like GRB\,051221A) and sub-energetic GRBs (like GRB\,060218) 
occupy the \emph{same} region of the $E_{\rm{pk}}$- $E_{\rm{\gamma,iso}}$-$E_{\rm{X,iso}}$ 
space. The same is true for the peculiar long GRB\,060614, 
later re-classified as a possible short GRB \citep{Gehrels06}.
In general, GRBs seem to divide into two groups with ``normal" long GRBs occupying the upper-right area;
short and peculiar GRBs together with XRFs (e.g. 050416A, 060218, 081007, 060614
also have a spectral peak energy  below 60 keV) share the same lower-left region of the plot.

\item The best-fitting slope of the $E_{\rm{X,iso}}$ vs. $E_{\rm{\gamma,iso}}$ relation of Fig. \ref{Fig:parcor2}
reads: $m=0.79\pm0.01$ (see Table \ref{tab_corr}). The significant departure of $m$ from 1 implies
the more energetic long GRBs to have a \emph{lower}   $\epsilon \equiv E_{\rm{X,iso}}/E_{\rm{\gamma,iso}}
\propto 1/E_{\rm{\gamma,iso}}^{0.2}$, with 
short GRBs being outliers of this relation.
Interestingly, equation \ref{Eq:3parcor} implies: $\epsilon\propto \frac{1}{E_{\rm{pk}}^{0.60}}$,
suggesting that the key parameter determining the $\gamma$-ray to X-ray ratio is not 
$E_{\rm{\gamma,iso}}$ but the \emph{spectral peak energy} $E_{\rm{pk}}$ \emph{irrespective of the nature of the GRB}
(either long or short). This is clear from the inset of Fig. \ref{Fig:3parcor}: the higher the prompt peak energy
the lower the $\epsilon$ (the GRB with $E_{\rm{pk}}\sim10^4$ keV is GRB\,090510). 
We refer to \cite{Zhang07} for a discussion of GRB
radiative efficiencies derived from X-ray data.
\item  The $\epsilon(E_{\rm{pk}})$ scaling above can be  interpreted as a \emph{physical} 
dependence of the radiative efficiency $\eta_{\gamma}$ on $E_{\rm{pk}}$:
$\eta_{\gamma}\equiv E_{\gamma}/(E_{\gamma}+E_{K})\approx E_{\gamma}/E_{K}\propto E_{\gamma}/E_{x}$
as long as $E_{\gamma}<E_{K}$. This would imply $\eta_{\gamma}\propto E_{\rm{pk}}^{0.6}$. See
\cite{Fan12} for a discussion  of this finding in the context of GRB photospheric models. 
Alternatively, a similar scaling could result
for the long population if long GRBs 
with lower \emph{isotropic} $E_{\rm{\gamma,iso}}$ are less beamed than high energy GRBs during the
prompt emission, but show otherwise similar beaming during the subsequent X-ray phase. In the first case,
the $\epsilon(E_{\rm{pk}})$ scaling would give direct information about the dissipative processes behind GRBs;
in the second case, it would be an observational effect, that nevertheless would provide valuable information about 
GRB jets and their opening angles. 
A complete and detailed discussion is beyond the scope of the present work and 
is provided by a companion paper (B12).
\item In Sec.  \ref{SubSec:ObsBias} we showed that 
the different time intervals over which $E_{\rm{X,iso}}$ has been estimated
do not severely affect  the $E_{\rm{X,iso}}$-$E_{\rm{\gamma,iso}}$- $E_{\rm{pk}}$.
\end{itemize}
\section{Summary and conclusions}
\label{Sec:conclusion}

We performed a comprehensive statistical analysis of Swift X-ray light-curves of 658 GRBs detected by XRT 
in the time period end of December 2004- end of December 2010. For the first time we present and analyse:
(i) the properties of GRBs in a common rest-frame 0.3-30 keV energy band; 
(ii) we furthermore perform a comparative study of long and short GRBs; (iv) we cross-correlate the
prompt $\gamma$-ray properties and the X-ray LCs properties. We report below a summary of the major 
findings.

From the \emph{spectral} analysis of GRBs with redshift (Sec. \ref{SubSubSec:nh}):
\begin{enumerate}
\item[1.] We find evidence for  high intrinsic neutral hydrogen absorption 
$\rm{NH_{HG}}\gtrsim 10^{22}\rm{cm^{-2}}$ even at $z\lesssim 2$. The average
value for long GRBs is  $\rm{NH_{HG}}\sim10^{21.9}\,\rm{cm^{-2}}$ (mean value of the
logarithm of the $\rm{NH_{HG}}$).
\item[2.] Short GRBs map the low end of the distribution with mean
$\rm{NH_{HG}}\sim10^{21.4}\,\rm{cm^{-2}}$. However, there is
\emph{no evidence} for short GRBs to show a lower  $\rm{NH_{HG}}$ when compared
to long GRBs in the \emph{same} redshift bin.  
\end{enumerate}

The analysis of  297 \emph{long} GRBs with complete X-ray light-curves\footnote{The total number of
C-like GRBs is 316 (Table \ref{Tab:lctype}). 19 are short GRBs.}
reveals that:
\begin{enumerate}
\item[3.] The average energy released in X-rays (0.3-30 keV, rest-frame)   
is $E_{\rm{X,iso}}\sim7\times 10^{51}$ erg typically representing $\sim 7\%$ of $E_{\rm{\gamma,iso}}$
(Sec. \ref{SubSubSec:Energetics}). 
The two quantities are statistically correlated: $E_{\rm{X,iso}}\propto E_{\rm{\gamma,iso}}^{0.8}$. 
Also: $E_{\rm{X,iso}}\propto E_{\rm{pk}}$ (Sec. \ref{SubSubSec:energycorr}).
\item[4.] The $E_{\rm{X,iso}}$ distribution does not extend beyond $10^{53}$ erg
(Sec. \ref{SubSubSec:Energetics}) possibly suggesting the
existence of a maximum available energy budget (the record holder is GRB\,080721 with 
$E_{\rm{X,iso}}\sim 10^{53}$ erg). Also, for $z>2$ we are not sensitive to the population of
GRBs with $E_{\rm{X,iso}}<10^{51}\,\rm{erg}$, so that the low-energy tail of 
the $E_{\rm{X,iso}}$ distribution is currently under-sampled.
\item[5.] The X-ray luminosity of the LCs at \emph{any} rest frame time between $100$ s and $10^5$ s is found to 
correlate with $E_{\rm{\gamma,iso}}$ (Sec \ref{SubSec:LEiso}): the scatter of this correlation increases with time
which might suggest that early time X-rays are more tightly related to the prompt phase.
\end{enumerate}

In the case of \emph{short} GRBs  (19 have C-like LCs):
\begin{enumerate}
\item[6.] The median luminosity light-curve of short GRBs (Sec. \ref{SubSec:aftdis}) is a factor $\sim10-30$ dimmer 
than long GRBs in the rest-frame time interval $10^2-10^4$ s, 
has a steeper average decay ($\propto t^{-1.3}$ vs. $\propto t^{-1}$) and shows
no evidence for clustering at late times (contrary to long GRBs).
\item[7.] Short GRBs populate the low-energy tail of the $E_{\rm{X,iso}}$ distribution, with 
$E_{\rm{X,iso}}^{short}\sim\frac{1}{50}E_{\rm{X,iso}}^{long}$ and an average
$E_{\rm{X,iso}}^{short}\sim10^{50}$ erg (Sec. \ref{SubSubSec:Energetics}).  
Short GRBs are more energy deficient during the second LC phase when compared
to long GRBs.
\item[8.] Short bursts are clear outliers of the $E_{\rm{X,iso}}-E_{\rm{\gamma,iso}}$ and 
$E_{\rm{X,iso}}-E_{\rm{pk}}$ relations established by the long population, with $E_{\rm{X,iso}}^{short}$
a factor $\gtrsim 50$ below expectations (Sec. \ref{SubSubSec:energycorr}). Short GRBs
are also found to lie below the $L_{\rm{X}}^{\rm{11hr}}$ vs. $E_{\rm{\gamma,iso}}$ relation 
established by the long class.
\item[9.] Short GRBs are under-represented in the class of GRBs showing clear evidence of
plateaus in the X-rays. Only 2 GRBs out of 19 possibly have plateaus (10\%). 
The corresponding percentage 
for long GRBs is instead $\sim 37\%$. While the limited sample size does not allow us to draw
firm conclusions, we note that  X-ray plateaus are more commonly detected in 
\emph{long} GRBs (Sec. \ref{SubSubSec:Lfenergy}).
\item[10.] The two short GRBs with X-ray LC plateaus in our sample are outliers of the  $L_{\rm{f}}$ vs.
$t_{\rm{f}}^{\rm{RF}}$ relation (Sec. \ref{SubSubSec:Lfenergy}). 
\end{enumerate}

Irrespective of the long or short GRB nature, we find no statistically significant correlation involving
the rest frame prompt duration $T_{90}^{RF}$, the intrinsic column density $\rm{NH}_{\rm{HG}}$
or the temporal slopes of the X-ray LCs (Sec. \ref{SubSec:parcor}).  The $T_{90}^{RF}$ basically
accounts for the second strongest Principal Component (Sec. \ref{SubSubSec:PCA}),
suggesting that while
$E_{\rm{pk}}$, $L_{\rm{pk}}$, $E_{\rm{X,iso}}$ and $E_{\rm{\gamma,iso}}$ are related to one another,
the $\gamma$-ray duration represents an \emph{additional} degree of freedom to the system.

We showed in Sec. \ref{SubSubSec:3parcor} the existence of a 3-parameter correlation that links
 $E_{\rm{X,iso}}$, $E_{\rm{\gamma,iso}}$ and $E_{\rm{pk}}$: 
 $E_{\rm{X,iso}}\propto (E_{\rm{\gamma,iso}}^{1.00}/E_{\rm{pk}}^{0.60})$:
\begin{enumerate}
\item Short and long GRBs share the \emph{same}  scaling.
\item This correlation implies $\frac{E_{\rm{\gamma,iso}}}{E_{\rm{X,iso}}}\propto E_{\rm{pk}}^{0.6}$
which can be interpreted as $\eta_{\gamma}\propto E_{\rm{pk}}^{0.6}$ (where $\eta_{\gamma}$
is the radiative efficiency).
\item Standard long GRBs and short GRBs (together with peculiar GRBs and XRFs) occupy a 
different region of the $E_{\rm{X,iso}}$-$E_{\rm{\gamma,iso}}$-$E_{\rm{pk}}$ plane.
\end{enumerate}

The results from our analysis are publicly available.\footnote{A demo
version of the website is currently available at 
http://www.grbtac.org/xrt\_demo/GRB060312Afterglow.html }  

\bibliographystyle{mn2e}
\bibliography{catalogo_articolo}
\section*{Acknowledgments}
RM thanks Lorenzo Amati and Lara Nava for sharing their data before publication. 
This research has made use of the XRT Data Analysis Software (XRTDAS) 
developed under the responsibility of the ASI Science Data Center (ASDC), Italy. 
MGB thanks ASI grant SWIFT I/004/11/0.
PAE, KLP and JPO acknowledge financial support from the UK Space Agency.
PR acknowledges financial contribution from the agreement ASI-INAF
I/009/10/0.

\appendix
\section{Glossary}
\label{Appendix:par}
This section provides the list of symbols used.
As a general note: 
X-ray energies (fluences) were computed from the time of the \emph{Swift}-XRT repointing up until the end of the 
observation; no temporal extrapolation was performed. The values reported
assume isotropic emission. X-ray fluences and fluxes are reported in the 0.3-10 keV (observer-frame) energy band;
energies, luminosities and intrinsic time scales are computed in the 0.3-30 keV (rest-frame) band.
\begin{itemize}
\item $\alpha_{\rm{n}}$: temporal slope of the normal decay phase.  Type Ia: $\alpha_{\rm{n}}=\alpha_2$;
type IIa: $\alpha_{\rm{n}}=\alpha_3$;  type III: $\alpha_{\rm{n}}=\alpha_4$.
\item $\alpha_{\rm{st}}$: temporal slope of the steep decay phase. Type Ib and IIa : $\alpha_{\rm{st}}=\alpha_1$; type IIb and III: 
$\alpha_{\rm{st}}=\alpha_2$ (see Fig. \ref{Fig:lctype}). The zero-time of the power-law decay is assumed to be the BAT
trigger time (i. e. $t_0=0$).
\item $\alpha_{\rm{st}}^{T90}$: temporal slope of the steep decay phase assuming $t_0=T_{90}$.
\item $\alpha_{\rm{sh}}$: temporal slope of the shallow decay (or plateau) phase. This corresponds to $\alpha_2$ and $\alpha_3$ for
type IIa and type III light-curves, respectively.
\item $\Gamma_x$: XRT 0.3-10 keV (observer frame) spectral photon index from this paper. 
\item $E_{\rm{\gamma,iso}}$: isotropic equivalent energy released during the prompt emission in the rest-frame $1-10^4$
keV energy band  from \cite{Amati08}.
\item $E_{\rm{pk}}$: rest-frame peak energy of the $\nu F_{\nu}$ spectrum during the prompt $\gamma$-ray emission from \cite{Amati08}.
\item $F_{\rm{f}}$ ($L_{\rm{f}}$): flux (luminosity) at the end of the plateau (i.e. at $t=t_{\rm{f}}$).
\item $F_{\rm{i}}$ ($L_{\rm{i}}$): flux (luminosity) at the beginning of the plateau (i.e. at $t=t_{\rm{i}}$).
\item $L_{\rm{pk,iso}}$: $1-10^4$ keV (rest frame) isotropic peak luminosity during the prompt emission from
\cite{Nava08}.
\item $L_{X}^{\rm{11h}}$:  luminosity at 11 hours rest-frame. 
\item $L_{X}^{\rm{10min}}$: luminosity at 10 min rest-frame. 
\item $\rm{NH_{\rm{tot}}}$: total  neutral hydrogen column density. 
\item $\rm{NH_{\rm{HG}}}$: intrinsic neutral hydrogen column density at the redshift of the GRB. 
\item $S_{\rm{1,X}}$ ($E_{\rm{1,X}}$): fluence (energy) released during the first phase of the X-ray light-curve.Type Ib and IIa: $E_{\rm{1,X}}=E_{\rm{1}}$; type IIb and III: $E_{\rm{1,X}}=E_{\rm{1}}+E_{\rm{2}}$. Fluences follow the same definition scheme. $E_{\rm{1}}$,  $E_{\rm{2}}$, $E_{\rm{3}}$ and $E_{\rm{4}}$ has been
defined following Fig. \ref{Fig:lctype}.
\item $S_{\rm{2,X}}$ ($E_{\rm{2,X}}$):  fluence (energy) released during the second phase of the X-ray light-curve.
Type Ia: $E_{\rm{2,X}}=E_{\rm{1}}+E_{\rm{2}}$; type Ib: $E_{\rm{2,X}}=E_{\rm{2}}$;
type IIa:  $E_{\rm{2,X}}=E_{\rm{2}}+E_{\rm{3}}$; type IIb: $E_{\rm{2,X}}=E_{\rm{3}}$; type III: $E_{\rm{2,X}}=E_{\rm{2}}+E_{\rm{4}}$ (see Fig. \ref{Fig:lctype}). Same definition scheme for fluences.  
\item $S_{\rm{\gamma}}$ ($E_{\rm{\gamma}}^{15-150}$): 15-150  keV (observer frame)  fluence (energy) released during the
prompt emission as calculated by \cite{Sakamoto11}.
\item $S_{\rm{X}}$ ($E_{\rm{X,iso}}$): X-ray fluence (energy).
\item $S_{\rm{X}}^{\rm{FL}}$ ($E_{\rm{X}}^{\rm{FL}}$): X-ray fluence (energy) associated to flares. For each GRB, the total fluence (energy) released in X-rays reads: $S_{\rm{X}}^{\rm{FL}}$+$S_{\rm{X}}$ ($E_{\rm{X}}^{\rm{FL}}$+$E_{\rm{X,iso}}$).
\item $S_{\rm{1,X}}^{\rm{FL}}$, $S_{\rm{2,X}}^{\rm{FL}}$ ($E_{\rm{1,X}}^{\rm{FL}}$, $E_{\rm{2,X}}^{\rm{FL}}$): X-ray fluence 
(energy) of flares superimposed on the first and second light-curve phase.
\item $t_{\rm{f}}$, $t_{\rm{f}}^{\rm{RF}}$, $t_{\rm{f}}^{\rm{T90}}$: end time of the plateau phase: observer frame, 
rest frame and in $T_{\rm{90}}$ units. 
This parameter corresponds to $t_{b2}$ and $t_{b3}$ for type IIa and type III light-curve, respectively.
\item $t_{\rm{i}}$, $t_{\rm{i}}^{\rm{RF}}$, $t_{\rm{i}}^{\rm{T90}}$: start time of the plateau phase: observer frame, rest frame and in $T_{\rm{90}}$
units. This parameter corresponds to $t_{b1}$ and $t_{b2}$ for type IIa and type III light-curve, respectively.
\item $\Delta t$: plateau duration defined as $t_{\rm{f}}-t_{\rm{i}}$.
\item $T_{90}$, $T_{90}^{\rm{RF}}$: duration of the 15-150 keV prompt emission from \cite{Sakamoto11}, in the observer and in the rest-frame, respectively.
\end{itemize}

\section{Tables}

\begin{table*}
\centering
\caption{Best-fitting parameters of the 0.3-10 keV (observer frame) light-curves in de-absorbed flux units, as obtained following the procedure outlined in Sec. \ref{SubSec:fitting}: GRB name, LC type (as defined in Sec. \ref{SubSec:fitting}), redshift, power-law indices ($\alpha_1$, $\alpha_2$, $\alpha_3$, $\alpha_4$) and errors, break times ($t_{b1}$, $t_{b2}$, $t_{b3}$) and errors, normalizations ($N_1$, $N_2$) and errors, smoothing parameters ($s_1$, $s_2$), prompt emission $T_{90}$ (we refer to \citealt{Sakamoto11} for GRBs detected before December 2009, and to the refined BAT GCNs otherwise), power-law index of the first segment when $t_0=T_{90}$ ($\alpha_1^{T90}$) and error, $\chi^2$, degrees of freedom, p-value.
Normalizations are given in $10^{-10}$ erg cm$^{-2}$ s$^{-1}$, times in seconds. A redshift equal to 0 indicates that no reliable estimate of this parameter is available from the literature. For the other columns, $-9$ indicates that the value is absent (i.e. there is no such LC phase). Note on the LC type column: C-GRBs (i.e. GRBs with complete LCs) are defined as promptly repointed GRBs $t_{\rm{rep}}<300$s whose fading was followed up to a factor 5-10 from the background limit; if this is not the case, the GRB is flagged as U-like (i.e. GRB with truncated LC). The flag F (N) indicates that flares have (have not) been detected. This table is available in its entirety in a machine-readable form in the online journal. A portion is shown here for guidance.}
\label{Table:bestfitpar}
\scalebox{0.9}{
\begin{tabular}{cccccccccccccccccccccccccc}
\hline
\hline
GRB & Type & $z$ & $\alpha_1$ & $\sigma_{\alpha1}$ & $\alpha_2$ & $\sigma_{\alpha2}$ & $\alpha_3$ & $\sigma_{\alpha3}$ & $\alpha_4$ & $\sigma_{\alpha4}$ & $Log[t_{b1}]$ & $\sigma_{tb1}$ & $Log[t_{b2}]$ & $\sigma_{tb2}$ & $Log[t_{b3}]$ & $\sigma_{tb3}$ \\
\hline
041223 & 0UN & $0$ & $1.91$ & $0.44$ & $-9.$ & $-9.$ & $-9.$ & $-9.$ & $-9.$ & $-9.$ & $-9.$ & $-9.$ & $-9.$ & $-9.$ & $-9.$ & $-9.$ \\
050124 & 0UN & $0$ & $1.44$ & $0.17$ & $-9.$ & $-9.$ & $-9.$ & $-9.$ & $-9.$ & $-9.$ & $-9.$ & $-9.$ & $-9.$ & $-9.$ & $-9.$ & $-9.$ \\
050126 & ICN & $1.29$ & $2.5$ & $0.51$ & $0.862$ & $0.25$ & $-9.$ & $-9.$ & $-9.$ & $-9.$ & $2.72$ & $0.4$ & $-9.$ & $-9.$ & $-9.$ & $-9.$ \\
050128 & IUN & $0$ & $0.758$ & $0.12$ & $1.38$ & $0.14$ & $-9.$ & $-9.$ & $-9.$ & $-9.$ & $3.69$ & $0.38$ & $-9.$ & $-9.$ & $-9.$ & $-9.$ \\
050219A & ICN & $0$ & $3.68$ & $0.36$ & $0.779$ & $0.11$ & $-9.$ & $-9.$ & $-9.$ & $-9.$ & $2.4$ & $0.054$ & $-9.$ & $-9.$ & $-9.$ & $-9.$ \\
050219B & 0UN & $0$ & $1.42$ & $0.044$ & $-9.$ & $-9.$ & $-9.$ & $-9.$ & $-9.$ & $-9.$ & $-9.$ & $-9.$ & $-9.$ & $-9.$ & $-9.$ & $-9.$ \\
050315 & IICN & $1.95$ & $-0.295$ & $0.18$ & $0.819$ & $0.037$ & $3.83$ & $0.18$ & $-9.$ & $-9.$ & $2.71$ & $0$ & $3.82$ & $0.14$ & $-9.$ & $-9.$ \\
\hline
\end{tabular}}
\end{table*}

\setcounter{table}{1}
\begin{table*}
\centering
\caption{Continued from Table~\ref{Table:bestfitpar}}
\scalebox{0.9}{
\begin{tabular}{cccccccccccc}
\hline
\hline
$N_1$ & $\sigma_{N1}$ & $N_2$ & $\sigma_{N2}$ & $s_1$ & $s_2$ & $T_{90}$ & $\alpha_1^{T90}$ & $\sigma_{\alpha T90}$ & $\chi^2$ & d.o.f. & p-val \\
\hline
$1.45e7$ & $6.3e7$ & $-9.$ & $-9.$ & $-9.$ & $-9.$ & $109.$ & $1.91$ & $0.44$ & $8.39$ & $25.$ & $0.999$ \\
$4.72e4$ & $8.e4$ & $-9.$ & $-9.$ & $-9.$ & $-9.$ & $3.93$ & $1.44$ & $0.17$ & $31.6$ & $30.$ & $0.384$ \\
$0.0673$ & $0.11$ & $-9.$ & $-9.$ & $0.5$ & $-9.$ & $48.$ & $2.5$ & $0.51$ & $3.54$ & $10.$ & $0.966$ \\
$0.458$ & $0.49$ & $-9.$ & $-9.$ & $-0.5$ & $-9.$ & $28.$ & $0.758$ & $0.12$ & $69.8$ & $148.$ & $1.$ \\
$0.457$ & $0.14$ & $-9.$ & $-9.$ & $1.$ & $-9.$ & $23.8$ & $3.13$ & $0.31$ & $56.6$ & $79.$ & $0.973$ \\
$1.26e5$ & $4.8e4$ & $-9.$ & $-9.$ & $-9.$ & $-9.$ & $28.7$ & $1.42$ & $0.044$ & $81.9$ & $127.$ & $0.999$ \\
$0.179$ & $0.027$ & $2.15e9$ & $1.8e9$ & $-0.5$ & $-9.$ & $95.6$ & $-0.295$ & $0.18$ & $155.$ & $218.$ & $1.$ \\
\hline
\end{tabular}}
\end{table*}

\begin{table*}
\centering
\caption{0.3-10 keV (observer-frame) fluence table. From left to right: GRB name, LC type (as defined in Sec. \ref{SubSec:fitting}), redshift, initial ($T_{min}$) and final ($T_{max}$) time of the observations, total fluence ($S_{X,iso}$) with error, fluence of the different LC phases ($S_1$, $S_2$, $S_3$, $S_4$) and errors, fluence of the flares in different parts of the LC ($S_1^{FL}$, $S_2^{FL}$, $S_3^{FL}$, $S_4^{FL}$) and errors. Fluences are given in erg cm$^{-2}$. A redshift equal to 0 indicates that no reliable estimate of this parameter is available from the literature. A ``$-9$" indicates that the LC does not show such phase and the value of that parameter is therefore absent.
Finally, for the columns containing information from flares superimposed on the power-law decay, 0 indicates that
no statistically significant positive fluctuation has been detected.
This table is available in its entirety in a machine-readable form in the online journal. A portion is shown here for guidance.}
\label{Table:bestfitene}
\scalebox{0.88}{
\begin{tabular}{cccccccccccccccccccccccccc}
\hline
\hline
GRB & Type & $z$ & $Log[T_{min}]$ & $Log[T_{max}]$ & $S_{X,iso}$ & $\sigma_{SX}$ & $S_1$ & $\sigma_{S1}$ & $S_2$ & $\sigma_{S2}$ & $S_3$ & $\sigma_{S3}$ & $S_4$ & $\sigma_{S4}$  \\
\hline
$041223$ & $0UN$ & $0$ & $4.22$ & $4.45$ & $8.8e-8$ & $6.7e-9$ & $-9.$ & $-9.$ & $-9.$ & $-9.$ & $-9.$ & $-9.$ & $-9.$ & $-9.$ \\
$050124$ & $0UN$ & $0$ & $4.05$ & $6.4$ & $1.6e-7$ & $3.4e-8$ & $-9.$ & $-9.$ & $-9.$ & $-9.$ & $-9.$ & $-9.$ & $-9.$ & $-9.$ \\
$050126$ & $ICN$ & $1.29$ & $2.12$ & $4.83$ & $3.7e-8$ & $6.8e-9$ & $1.5e-8$ & $2.e-9$ & $2.2e-8$ & $6.3e-9$ & $-9.$ & $-9.$ & $-9.$ & $-9.$ \\
$050128$ & $IUN$ & $0$ & $2.23$ & $4.84$ & $7.9e-7$ & $2.4e-8$ & $4.6e-7$ & $1.8e-8$ & $3.3e-7$ & $1.5e-8$ & $-9.$ & $-9.$ & $-9.$ & $-9.$ \\
$050219A$ & $IUN$ & $0$ & $2.05$ & $6.2$ & $3.5e-7$ & $1.4e-7$ & $4.1e-8$ & $2.e-9$ & $3.1e-7$ & $1.4e-7$ & $-9.$ & $-9.$ & $-9.$ & $-9.$ \\
$050219B$ & $0UN$ & $0$ & $3.5$ & $6.22$ & $9.4e-7$ & $6.1e-8$ & $-9.$ & $-9.$ & $-9.$ & $-9.$ & $-9.$ & $-9.$ & $-9.$ & $-9.$ \\
$050315$ & $IICN$ & $1.95$ & $1.92$ & $5.93$ & $1.1e-6$ & $3.3e-8$ & $2.4e-7$ & $1.4e-8$ & $7.e-8$ & $8.8e-9$ & $8.e-7$ & $2.8e-8$ & $-9.$ & $-9.$ \\
\hline
\end{tabular}}
\end{table*}

\setcounter{table}{3}
\begin{table*}
\centering
\caption{Continued from Table~\ref{Table:bestfitene}.}
\scalebox{0.9}{
\begin{tabular}{cccccccccccccccccccccccccc}
\hline
\hline
$S_1^{FL}$ & $\sigma_{S1^{FL}}$ & $S_2^{FL}$ & $\sigma_{S2^{FL}}$ & $S_3^{FL}$ & $\sigma_{S3^{FL}}$ & $S_4^{FL}$ & $\sigma_{S4^{FL}}$  \\
\hline

$0$ & $0$ & $-9.$ & $-9.$ & $-9.$ & $-9.$ & $-9.$ & $-9.$ \\
$0$ & $0$ & $-9.$ & $-9.$ & $-9.$ & $-9.$ & $-9.$ & $-9.$ \\
$0$ & $0$ & $0$ & $0$ & $-9.$ & $-9.$ & $-9.$ & $-9.$ \\
$0$ & $0$ & $0$ & $0$ & $-9.$ & $-9.$ & $-9.$ & $-9.$ \\
$0$ & $0$ & $0$ & $0$ & $-9.$ & $-9.$ & $-9.$ & $-9.$ \\
$0$ & $0$ & $-9.$ & $-9.$ & $-9.$ & $-9.$ & $-9.$ & $-9.$ \\
$0$ & $0$ & $0$ & $0$ & $0$ & $0$ & $-9.$ & $-9.$ \\

\hline
\end{tabular}}
\end{table*}

\begin{table*}
\centering
\caption{0.3-30 keV (rest-frame) energy table. From left to right: GRB name, LC type (as defined in Sec. \ref{SubSec:fitting}), redshift, initial ($T_{min}$) and final ($T_{max}$) time of the observations, total energy ($E_{X,iso}$) with error, energy of the different LC phases ($E_1$, $E_2$, $E_3$, $E_4$) and errors, energy of the flares in different parts of the LC ($E_1^{FL}$, $E_2^{FL}$, $E_3^{FL}$, $E_4^{FL}$) and errors. Energies are given in erg. A redshift equal to 0 indicates that no reliable estimate of this parameter is available from the literature. A ``$-9$" indicates that the LC does not show such phase and the value of that parameter is therefore absent.
Finally, for the columns containing information from flares superimposed on the power-law decay, 0 indicates that
no statistically significant positive fluctuation has been detected.
This table is available in its entirety in a machine-readable form in the online journal. A portion is shown here for guidance regarding its form and content.}
\label{Table:bestfitene0330}
\scalebox{0.88}{
\begin{tabular}{cccccccccccccccccccccccccc}
\hline
\hline
GRB & Type & $z$ & $Log[T_{min}]$ & $Log[T_{max}]$ & $E_{X,iso}$ & $\sigma_{EX}$ & $E_1$ & $\sigma_{E1}$ & $E_2$ & $\sigma_{E2}$ & $E_3$ & $\sigma_{E3}$ & $E_4$ & $\sigma_{E4}$  \\
\hline
$050126$ & $\rm{ICN}$ & $1.29$ & $2.12$ & $4.83$ & $3.7e-8$ & $6.8e-9$ & $1.5e-8$ & $2.e-9$ & $2.2e-8$ & $6.3e-9$ & $-9.$ & $-9.$ & $-9.$ & $-9.$ \\
$050315$ & $\rm{IICN}$ & $1.95$ & $1.92$ & $5.93$ & $1.1e-6$ & $3.3e-8$ & $2.4e-7$ & $1.4e-8$ & $7.e-8$ & $8.8e-9$ & $8.e-7$ & $2.8e-8$ & $-9.$ & $-9.$ \\
$050318$ & $\rm{0UN}$ & $1.44$ & $3.52$ & $5.65$ & $2.5e-7$ & $1.1e-8$ & $-9.$ & $-9.$ & $-9.$ & $-9.$ & $-9.$ & $-9.$ & $-9.$ & $-9.$ \\
$050319$ & $\rm{IICN}$ & $3.24$ & $1.98$ & $6.15$ & $5.e-7$ & $5.e-8$ & $2.5e-8$ & $3.6e-9$ & $1.7e-7$ & $6.7e-9$ & $3.e-7$ & $4.9e-8$ & $-9.$ & $-9.$ \\
$050401$ & $\rm{ICF}$ & $2.9$ & $2.14$ & $5.9$ & $1.2e-6$ & $6.2e-8$ & $5.e-7$ & $7.5e-9$ & $7.3e-7$ & $6.1e-8$ & $-9.$ & $-9.$ & $-9.$ & $-9.$ \\
$050408$ & $\rm{0UN}$ & $1.24$ & $3.41$ & $6.47$ & $5.4e-7$ & $3.9e-8$ & $-9.$ & $-9.$ & $-9.$ & $-9.$ & $-9.$ & $-9.$ & $-9.$ & $-9.$ \\
\hline
\end{tabular}}
\end{table*}

\setcounter{table}{5}
\begin{table*}
\centering
\caption{Continued from Table~\ref{Table:bestfitene0330}.}
\scalebox{0.9}{
\begin{tabular}{cccccccccccccccccccccccccc}
\hline
\hline
$E_1^{FL}$ & $\sigma_{E1^{FL}}$ & $E_2^{FL}$ & $\sigma_{E2^{FL}}$ & $E_3^{FL}$ & $\sigma_{E3^{FL}}$ & $E_4^{FL}$ & $\sigma_{E4^{FL}}$  \\
\hline
$0$ & $0$ & $0$ & $0$ & $-9.$ & $-9.$ & $-9.$ & $-9.$ \\
$0$ & $0$ & $0$ & $0$ & $0$ & $0$ & $-9.$ & $-9.$ \\
$0$ & $0$ & $-9.$ & $-9.$ & $-9.$ & $-9.$ & $-9.$ & $-9.$ \\
$0$ & $0$ & $0$ & $0$ & $0$ & $0$ & $-9.$ & $-9.$ \\
$0$ & $0$ & $0$ & $0$ & $-9.$ & $-9.$ & $-9.$ & $-9.$ \\
$0$ & $0$ & $-9.$ & $-9.$ & $-9.$ & $-9.$ & $-9.$ & $-9.$ \\

\hline
\end{tabular}}
\end{table*}

\label{lastpage}
\end{document}